\documentclass[preprint]{elsarticle}

\usepackage{epsfig}
\usepackage{graphicx}
\usepackage{amssymb}
\usepackage{algorithm}
\usepackage{moreverb}
\usepackage{multirow}
\usepackage{rotating}
\usepackage{subfigure}

\usepackage[noend]{algcompatible}

\algblockdefx{FORALLP}{ENDFAP}[1]%
  {\textbf{for all }#1 \textbf{ in parallel}}

\usepackage{color}
\usepackage{listings}
\begin{document}

\title{Region Templates: Data Representation and Management for Large-Scale Image Analysis}

\author{George Teodoro\corref{fn1}$^3$}
\ead{teodoro@cic.unb.br}
\author{Tony Pan$^1$}
\ead{tony.pan@emory.edu}
\author{Tahsin Kurc$^{1,2}$}
\ead{tkurc@emory.edu} 
\author{Jun Kong$^1$}
\ead{jun.kong@emory.edu}
\author{Lee Cooper$^1$}
\ead{lee.cooper@emory.edu}
\author{Scott Klasky$^2$}
\ead{klasky@ornl.org}
\author{Joel Saltz$^1$}
\ead{jhsaltz@emory.edu}
\address{$^1$Biomedical Informatics Department, Emory University, Atlanta, GA, USA \\
$^2$Scientific Data Group, Oak Ridge National Laboratory, Oak Ridge, TN, USA\\
$^3$Department of Computer Science, University of Bras\'{\i}lia, Bras\'{\i}lia, DF, Brazil}

\cortext[fn1]{Corresponding author}
\begin{abstract}
Distributed memory machines equipped with CPUs and GPUs (hybrid computing
nodes) are hard to program because of the multiple layers of memory and
heterogeneous computing configurations. In this paper, we introduce a region
template abstraction for the efficient management of common data types used in
analysis of large datasets of high resolution images on clusters of hybrid
computing nodes. The region template provides a generic container template for
common data structures, such as points, arrays, regions, and object sets,
within a spatial and temporal bounding box. The region template abstraction
enables different data management strategies and data I/O implementations,
while providing a homogeneous, unified interface to the application for data
storage and retrieval.  The execution of region templates applications is
coordinated by a runtime system that supports efficient execution in hybrid
machines. Region templates applications are represented as hierarchical
dataflow in which each computing stage may be represented as another dataflow
of finer-grain tasks. A number of optimizations for hybrid machines are
available in our runtime system, including performance-aware scheduling for
maximizing utilization of computing devices and techniques to reduce impact of
data transfers between CPUs and GPUs.  An experimental evaluation on a
state-of-the-art hybrid cluster using a microscopy imaging study shows that
this abstraction adds negligible overhead (about 3\%) and achieves good
scalability. The implementations of global region templates storage achieved
very high data transfer throughput (200GB/s). The optimizations proposed in the
high speed disk based storage implementation to support asynchronous
applications resulted in an application performance gain of about 1.13$\times$
as compared to a state-of-the-art system.  Finally, the optimized cooperative
execution in a cluster with 100 nodes (300~GPUs and 1,200~CPU cores) attained a
processing rate of 11,730 4K$\times$4K tiles per minute. This computation rate
is enabling much larger studies, which we expect will also result into
significant scientific discoveries.

\end{abstract}


\maketitle

\section{Introduction} \label{sec:intro}
Distributed-memory computing systems consisting of multi-core CPUs and general purpose Graphics 
Processing Units (GPUs) provide large memory space and processing capacity for scientific 
computations. Leveraging these hybrid systems, however, is challenging 
because of multiple memory hierarchies and the different computation characteristics 
of CPUs and GPUs. Application developers have to deal with efficiently mapping and 
scheduling analysis operations onto multiple computing nodes and 
CPU cores and GPUs within a node, while enforcing dependencies between 
operations. They also need to implement mechanisms to stage, distribute, 
and manage large volumes of data and large numbers of data elements 
across memory/storage hierarchies. 

We have developed an analytics framework that can be used by application 
developers and users for on-demand, high throughput processing of very large 
microscopy image datasets on a hybrid computation 
system~\cite{Teodoro-IPDPS2013,Teodoro-IPDPS2012}.
The analytics framework consists of a library of high performance image analysis 
methods, data structures and methods common in microscopy image analyses, 
and a middleware system. In earlier work~\cite{Teodoro-IPDPS2013,Teodoro-IPDPS2012}, 
we presented methods and their implementations at the middleware layer for 
scheduling data analysis operations and two-level analysis pipelines the computation 
system. In this paper, we investigate efficient data representations and the associated 
runtime support to minimize data management overheads of common data types consumed and 
produced in the analysis pipeline.

The primary motivation for this effort is the quantitative 
characterization of disease morphology at the sub-cellular scale using 
large numbers of whole slide tissue images (WSIs). This is an important and 
computationally expensive application in biomedical research. 
Investigations of tissue morphology using WSI data (also referred to here 
as microscopy image 
data) have huge potential to lead to a much more sophisticated understanding of 
disease subtypes and feature distributions and enable novel methods for 
classification of disease state.
Biomedical researchers are now able to capture 
a highly detailed image of a whole slide tissue in a few minutes using 
state-of-the-art microscopy scanners. These devices are becoming more 
widely available at lower price points, making it feasible for research groups 
and organizations to collect large numbers of whole slide tissue images (WSIs) 
in human and animal studies – The Cancer Genome Atlas project, for instance, has more 
than 20,000 WSIs. We expect that in the next 3-5 years, research groups will 
be able to collect tens of thousands of digital microscopy images per study, 
and healthcare institutions will have repositories containing millions of images.

Over the past several years, a number of research groups, including our own, 
have developed and demonstrated a rich set of methods 
for carrying out quantitative microscopy image analyses and their applications 
in research~\cite{HAN12,PHAN12,KOTHARI12,CHIELA12,MONACO08,GURCAN06,COOPER10,jamia-insilico}.
Scaling such analyses to large numbers of images (and patients) 
creates high end computing and big data challenges. A typical analysis of a 
single image of $10^5$x$10^5$ pixel resolution involves extraction of millions 
of micro-anatomic structures and computation of 10-100 features per structure. 
This process may take several hours on a workstation. 

Our earlier work has demonstrated that large numbers of images can be processed rapidly 
using distributed memory hybrid systems by carefully scheduling analysis operations across 
and within nodes in the system and that scheduling decisions can be pushed to the 
middleware layer, relieving the application developer of implementing complex, 
application-specific scheduling mechanisms. The work presented in this paper introduces an 
abstraction layer, referred to here as the region template framework, for management and 
staging of data during the execution of an analysis application and shows that 
the overhead of such an abstraction is small. Our contributions can be 
summarized as follows: 
\begin{itemize}
\item A novel region template abstraction
to minimize data management overheads of common data types in large scale WSI analysis. 
The region template provides a generic container template for common data structures, such
as points, arrays, regions, and object sets, within a spatial and temporal
bounding box. A \emph{data region} object is a storage materialization of data
types and stores the data elements in the region contained by a region template
instance. A region template instance may have multiple data regions. 
The region template abstraction allows for different data I/O, storage, and management 
strategies and implementations, while providing a homogeneous, unified interface 
to the application developer.
\item An efficient runtime middleware to support the definition, materialization, and 
management of region templates and data regions and execution of analysis pipelines 
using region templates on distributed-memory hybrid machines. 
Application operations interact with data regions and region templates to store
and retrieve data elements, rather than explicitly handling the management,
staging, and distribution of the data elements. This responsibility is pushed
to the runtime system. Presently, the runtime system has implementations for 
memory storage on nodes with multi-core CPUs and GPUs, distributed memory storage, and
high bandwidth disk I/O.  
\item An experimental evaluation of the region template framework on a distributed
memory cluster system in which each compute node has 2 6-core CPUs and 3 NVIDIA
GPUs. The results demonstrate that this level of abstraction is highly scalable
and adds negligible overhead (about 3\%).
\end{itemize} 

While our target in this paper is large-scale microscopy analysis, there is a
broader class of applications that have similar data processing patterns and
employ similar data structures (e.g., arrays, regions, and object sets within a
spatial and temporal bounding box). This class includes applications for
satellite data processing, subsurface and reservoir characterization, and
analysis of astronomy telescope
datasets~\cite{DBLP:journals/ewc/KlieBGWSSPCSK06,DBLP:journals/concurrency/KurcCZSMWPSHSSSTP05}
~\cite{DBLP:conf/iccS/ParasharMBKRKCSW05,beynon01datacutter,Shock199865,Chang:1997:THR:645482.653300}.
Thus, we expect that our work will be applicable in other application domains.

The rest of the paper is organized as follows. Section~\ref{sec:back}
provides an overview of the motivating scenario, and the use-case application
for integrative microscopy image analysis. The region template framework is
described in Section~\ref{sec:rt}. Implementation of global region templates
data storage, which are used for inter-machine data exchange is discussed in
Section~\ref{sec:storage-impl}. Section~\ref{sec:res} presents an experimental
performance evaluation of the region template framework. The related work and
conclusions are presented in Sections~\ref{sec:related}~and~\ref{sec:conc}.

\section{Background}
\label{sec:back}

\subsection{Motivation}
While our work is primarily motivated by studies that analyze and integrate 
morphological information from tissue specimens, these studies belong to a 
larger class of scientific applications that analyze and mine low-dimensional, 
spatio-temporal data. These datasets are characterized by elements defined as 
points in a multi-dimensional coordinate space with low dimensionality and at 
multiple time steps. A point is primarily connected to points in its spatial 
neighborhood. These properties are common in many datasets from sensors and 
generated from scientific simulations: satellite data in climate studies, 
seismic surveys and numerical simulations in subsurface characterization, and 
astronomical data from telescopes. 

Rapid processing of data from remote sensors attached to earth orbiting satellites, 
for example, is necessary in weather forecasting and disaster tracking applications 
as well as for studying changes in vegetation and ocean ecosystems. Accurate prediction 
of weather patterns using satellite sensor data can assist a researcher to estimate 
where tornadoes may occur and their paths or where flooding because of heavy rain 
may take place. In this scenario, it is critical to analyze large volumes of data and 
corroborate the analysis results using multiple, complementary datasets (e.g., multiple 
types of satellite imagery). A dataset in this scenario may define regions of regular, 
lower resolution grids over the entire continent, while another may contain sensor 
readings on a higher resolution grid corresponding to a smaller spatial region. The 
researcher may perform a series of operations to (1) remove anomalous measurements and 
convert spectral intensities to value of interest; (2) map data elements in one dataset 
to another to create a mosaic of regions for full sensor coverage; (3) segment and 
classify regions with similar surface temperature; and (4) perform time-series 
calculations on land and air conditions; and (5) perform comparisons of land and air 
conditions over multiple time steps and across spatial regions to look for 
changing weather patterns. 

In subsurface characterization, as another example, scientists carry out simulations of 
complex numerical models on unstructured, multi-resolution meshes, which represent 
underground reservoirs being studied, to investigate long 
term changes in reservoir characteristics and examine what-if scenarios (e.g., for 
maximizing oil production by different placements of injection and production wells). 
Data are also obtained via multiple seismic surveys of the target region, albeit at 
lower spatial resolutions. A researcher may process, combine, and mine simulation and 
seismic datasets through a series of operations, including (1) selection of regions 
of interest from a larger spatio-temporal region; (2) mapping data elements from 
different datasets to a common coordinate system and resolution for analysis; (3) 
detecting, segmenting, and classifying pockets of subsurface materials (e.g., oil); 
(4) analyzing changes in segmented objects over time or under different initial conditions; 
and (5) correlating information obtained from one dataset with information obtained from 
another dataset (e.g., comparing segmented pockets from simulation datasets with those from 
seismic datasets to validate simulations).


\subsection{Use Case: Integrative Microscopy Image Analysis} \label{sec:motivating}
%

Microscopic examination of biopsied tissue reveals visual morphological
information enabling the pathologist to render accurate diagnoses, assess
response and guide therapy. Whole slide digital imaging enables this process to
be performed using digital images instead of physical slides. The quantitative
characterization of microscopy image data involves a process
of~\cite{jamia-insilico,kong2013machine,kong2013novel}: (1) correcting for staining and imaging artifacts,
(2) detection and extraction of microanatomic objects, such as nuclei and
cells, (3) computing and characterizing their morphologic and molecular
features, and (4) monitoring and quantifying changes over space and time. In
some imaging studies, processing also includes 3-D and/or spatio-temporal
reconstruction.  

\begin{figure*}[htb!]
\begin{center}
\includegraphics[width=\textwidth]{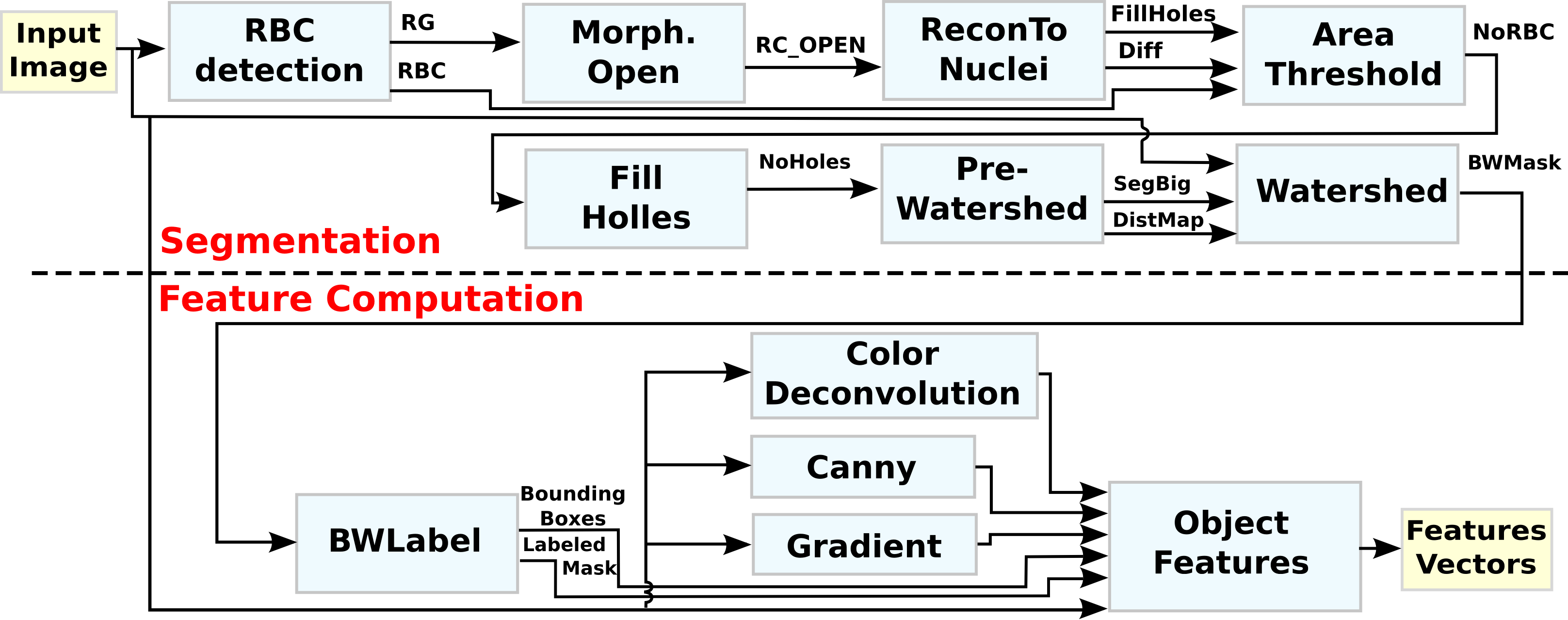}
\vspace*{-1ex}
\caption{An example pipeline of the segmentation and feature computation
stages. The segmentation phase identifies nuclei and cells in the input images, and defines
a cytoplasmic region around each nucleus. The feature computation stage calculates
a vector of 60-100 shape and texture features for each nucleus and cytoplasm found. Each
stage is internally described as a complex workflow of fine-grain tasks.}
\vspace*{-2ex}
\label{fig:app}
\end{center}
\end{figure*}

In a typical analysis scenario, nuclei and cells are first segmented in each
image, and a cytoplasmic space is defined around each nucleus in {\em the
segmentation stage}.  Features are calculated for each nucleus to describe its
shape and texture in {\em the feature computation stage}.  The properties of
the "average" nucleus for each patient is calculated to generate a patient
morphology profile. The patient morphology profiles are clustered using a
machine-learning algorithm in {\em the classification stage}.  The data are
normalized and redundant features are eliminated using a feature selection
process. The selected features are processed via statistical and machine
learning algorithms to group patients into clusters with cohesive morphological
characteristics.  The patient clustering results are further processed to
search for significant associations with clinical and genomics information in
{\em the correlation stage}.  The clusters are checked for differences in
patient outcome, associations with the molecular subtypes defined in
literature, human descriptions of pathologic criteria, and recognized genetic
alterations. 

In our current studies the most time consuming stages are the 
segmentation and feature computation stages. It is highly desirable 
in research studies to use large datasets in order to obtain robust, 
statistically significant results, but the scale of an image-based 
study is often limited by the computational cost of these stages. 
Modern scanners can generate a whole slide tissue
image at up to 120Kx120K-pixel resolutions. An uncompressed, 4-channel
representation of such an image is about 50GB.  Image analysis
algorithms segment $10^5$ to $10^7$ cells and nuclei in each virtual
slide of size $10^5$ by $10^5$ pixels. For each segmented object
50-100 shape and texture features are computed in the feature
computation phase.  Figure~\ref{fig:app} presents the computation
graph for the segmentation and features computation stages. 
The graph of operations performed within
each of these two stages is detailed. Processing a few hundred large 
resolution images on a workstation may take days. Distributed 
memory clusters of multi-core CPUs and modern GPUs
can provide the computational capacity and memory space needed for
analyses involving large numbers of images.

\section{Region Templates} \label{sec:rt}

\subsection{Region Templates Framework Architecture}

\begin{figure}[htb!]
\begin{center}
\includegraphics[width=0.95\textwidth]{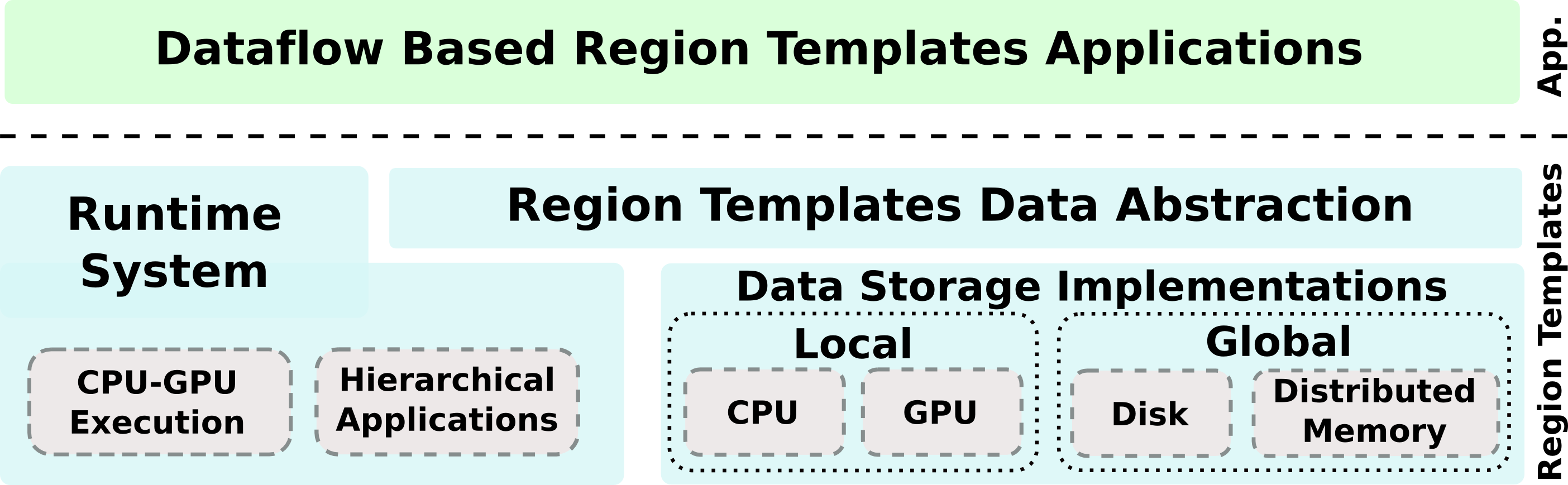}
\vspace*{-2ex}
\caption{Architecture of the Region Templates framework: Region Templates
supports representation and execution of applications using a hierarchical
dataflow model with support for hybrid systems, equipped with CPUs and GPUs.
The information read from/to application dataflow components are expressed
using region templates data abstraction, which includes a template for data
types commonly used in applications that process datasets represented in a
spatial and temporal domain. Multiple implementations of the data abstractions
are provided. These implementations include node local data instances stored in
GPU and CPU memory, as well as global storage that makes region templates data
regions visible for application components running in different nodes.  Global
data regions are used, i.e., to exchange data among application stages. The
current implementations of the Global Data Storage include: (i)~the high
performance disk storage approach that is primarily used to efficient stage
persistent data to disk; and, (ii)~the distributed memory based storage that
provides an efficient and transparent mechanism to transfers data among
components of the application.}
\vspace*{-2ex}
\label{fig:rt-arch}
\end{center}
\end{figure}

This section presents an overview of the Region Templates framework.  The main
modules of the system are depicted in Figure~\ref{fig:rt-arch}: runtime system,
region templates data abstraction, and implementations of data storage.

Region Templates applications are represented and executed as dataflows. It is
responsibility of \emph{the runtime system} module to manage execution on
distributed environments. The runtime system instantiates the application
components/stages, asserts dependencies among component instances are respected, optimizes
execution on hybrid CPU-GPU equipped systems, and performs a multi-level task
scheduling. Our runtime system implements a special type of dataflows
representation called hierarchical dataflow. This model allows for an
application to be described as a dataflow of coarse-grained components in which
each component may be further implemented as a dataflow of fine-grained tasks.
This representation leads to flexibility and improved scheduling performance on
hybrid systems, as detailed in Section~\ref{sec:runtime},

The data consumed and produced by the application components are stored using
containers provided by the \emph{Region Templates Data Abstraction}. The data types
available with this abstraction include those that are commonly found in
applications that process data represented in low-dimensional spaces (1D, 2D or
3D space) with a temporal component. Namely, some of the supported types are
pixels, points, arrays, matrices, 3D volumes, objects and regions of interest
that may be represented using polygons.

The region templates data abstraction implements efficient data transport
mechanisms for moving data among region templates application components, which
may run on different nodes of a distributed memory machine. Region templates
exchanged among application coarse-grained stages are called global. Instead of
writing data through streams as in a typical dataflow application, the
components of a region templates application output region templates data
instances (data regions) that are consumed by other components of the
applications. The dependencies among component instances and data regions
inputted/outputted by stages are provided to the runtime system by the
application. It is responsibility of the runtime system to coordinate the
transfers of data among stages. Data transfers are performed in background to
useful computation by I/O threads, which interact with the appropriate
implementations of the global data storage to retrieve/stage data. After
transfers are done, the user component code is launched to perform the intended
data transformations according to runtime system scheduling policies.

We currently provide implementations for global region templates storage using
high performance disks and distributed shared memory. We want to highlight that
both implementations should coexist and that they are not necessarily
competitors.  There are cases in which the use of a high performance disk based
mechanism is desirable, for instance, if the application needs to persist data
exchanged among components for further analysis. If the goal is to transfer
data as quickly as possible among application stages, using the distributed
shared memory implementation tend to be the best choice.  As such, we want to
provide multiple global storage mechanisms, which could be interchanged by the
end-user with minimum effort.

The description of the Region Templates framework is organized as follows:
Section~\ref{sec:runtime} presents the runtime system and the extensions
implemented to execute region templates applications;
Section~\ref{sec:rt-data-abstraction} presents the region templates data
abstraction; and, Section~\ref{sec:compose} describes the application
composition process and interaction among region templates components in the
execution. The storage implementations for region templates data abstractions
are discussed in Section~\ref{sec:storage-impl}.

\subsection{Runtime System Support}
\label{sec:runtime}

The runtime system used to coordinate the execution of region templates
applications is built on top of our previous
works~\cite{Teodoro-IPDPS2013,Teodoro-IPDPS2012}. In this section, we present
the core features of the runtime system that are inherited by region templates
applications, as well as the extensions implemented to handle execution of
region templates applications.

\begin{figure}
\begin{center}
\includegraphics[width=0.95\textwidth]{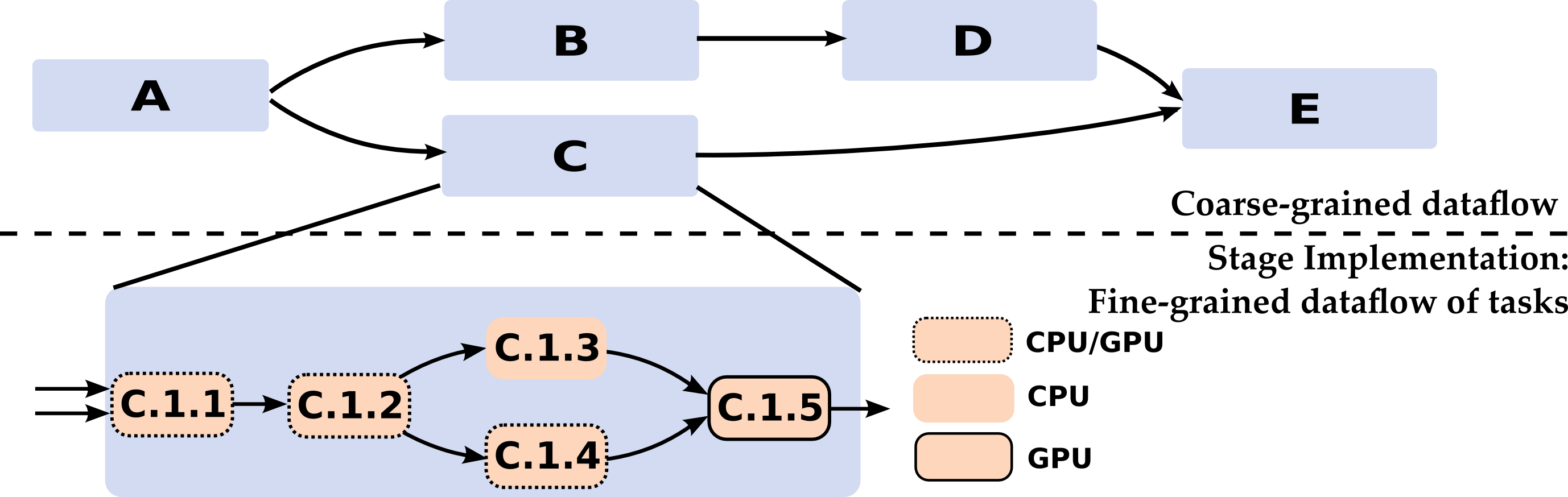}
\vspace*{-1ex}
\caption{Hierarchical dataflow model.  Each stage of an analysis pipeline may
be expressed as another graph of fine-grain operations. This results in a
hierarchical (two-level) computation graph. During execution stages are
are mapped to a computation node and fine-grain operations are dispatched as
tasks and scheduled for execution with CPUs and GPUs on that node.}
\vspace*{-2ex}
\label{fig:dataflow-representation}
\end{center}
\end{figure}

The application representation supported in the runtime system draws from
filter-stream model implemented in
DataCutter~\cite{beynon01datacutter,10.1109/ICPP.2008.72}. Filter-stream
applications are decomposed into components, connected to each other through
logical streams; a component reads data from one or more streams, computes the data
transformations, and writes the results to one or more streams. This model was
adapted in our system in the following ways.  This new framework supports
hierarchical dataflow in that an operation can itself be made up of a dataflow
of lower-level operations. It is described in the context of two dataflow
levels for sake of presentation, but it allows for multiple levels of
hierarchies. The first level is the coarse-grain operations level, which
represents the main stages of an analysis application. The fine-grain
operations level is the second level and represents lower-level operations,
from which a main stage is created.  Figure~\ref{fig:dataflow-representation}
illustrates the hierarchical pipeline representation of an analysis
application. 

The hierarchical dataflow representation allows for different scheduling
strategies to be used at each level. Fine-grain tasks can be dispatched for
execution with a scheduler in each node of the system, which is more flexible
then describing each dataflow component as a single task that should be
completely executed using a single device. With this strategy, it is possible
to exploit performance variability across fine-grain tasks and to better use
available devices.

\begin{figure}[htb!]
\begin{center}
\includegraphics[width=0.85\textwidth]{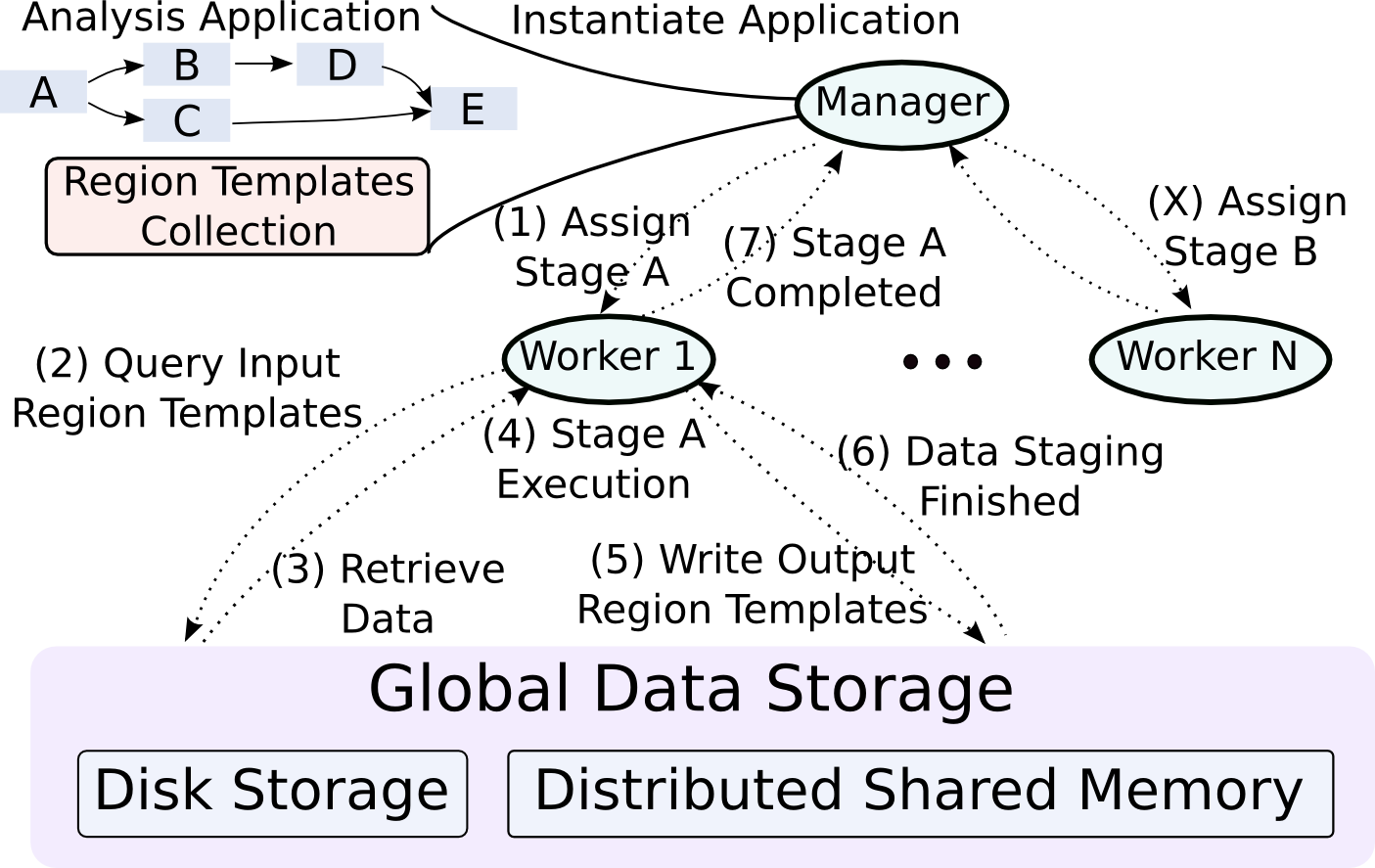}
\vspace*{-2ex}
\caption{Overview of the framework execution model and architecture. The
execution strategy of the system is built on top of a Manager-Worker model,
which executes the coarse-grain dataflow model in a bag-of-tasks style. The
application developer implements a part of the Manager module that
instantiates the application workflow: creating as many instances of each stage as
necessary and setting dependencies among them. At runtime, the Manager
assigns stage instances for computation with Worker nodes in a demand-driven
basis. Multiple stage instances may be active with a single Worker, which
communicates with the global data region storage to retrieve data used by each
stage instance. After a stage instance execution is completed, its output data
regions are written to the adequate implementation of the global data storage,
and Manager is notified. The Manager then releases appropriate dependencies,
and may dispatch other stage instances for execution.  }
\vspace*{-2ex}
\label{fig:execModelRuntime}
\end{center}
\end{figure}

The implementation of our runtime system is constructed using a Manager-Worker
model, which combines a bag-of-tasks execution with the coarse-grain dataflow
pattern. The application Manager creates instances of the coarse-grain stages
that include \emph{input data regions}, and exports the dependencies among the
stage instances. The dependency graph is not necessarily known prior to
execution, but may be built incrementally at runtime as a stage may create
other stage instances. The assignment of work from the Manager to Worker nodes
is performed at the granularity of a stage instance. The Manager schedules
stage instances to Worker in a demand-driven basis, and Workers repeatedly
request work until all applications stage instances have been executed (See
Figure~\ref{fig:execModelRuntime}).  Each Worker may execute several stage
instances concurrently in order to take advantage of multiple computing devices
available in a node. The communication between Manager and Worker components of
the runtime system is implemented using MPI. Data are read/written by stage
components using global data regions, which are implemented as a module of the
region templates framework and is responsible for inter-stage communication.
Once a stage instance is received, the WCT identifies all region templates used
by that stage, allocates memory locally on that node to store the regions, and
communicates with the appropriate storage implementation to retrieve those
regions. Only after data is ready in the node local memory, the stage instance
may start executing. The process of reading data overlaps with useful
computation, since tasks created by other stage instances may be concurrently
executing with data movement of current stage.

\begin{figure*}[htb!]
\begin{center}
\includegraphics[width=1\textwidth]{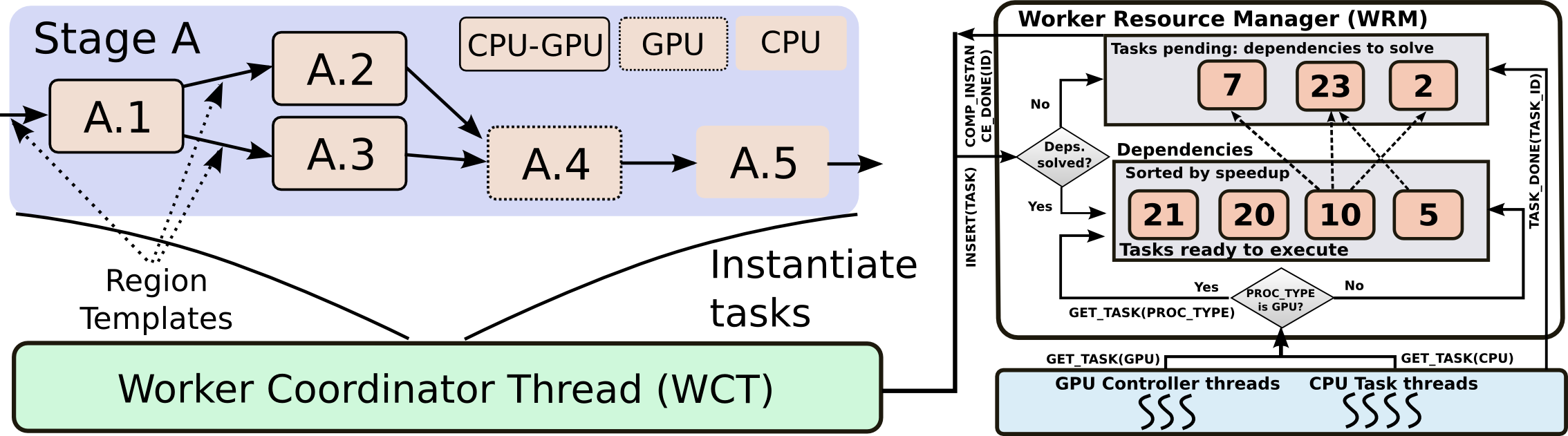}
\vspace*{-1ex}
\caption{As described, each stage of an analysis application may be expressed
as another graph of finer grain functions. Functions within a stage instance
mapped to a computation node are dispatched as tasks and scheduled for
execution by the Worker Resource Manager (WRM) on that node. The WRM creates
one computing threads to manage each CPU core or coprocessor, and assigns tasks
for concurrent execution in the available devices. Tasks ready for execution
are placed into a queue, sorted by speedup if our performance aware scheduler
described in Section~\ref{sec:opts} is used, and tasks with dependencies to be
resolved stay on a list of pending tasks. As the computing threads become idle,
a task ready to execute is assigned to that thread according to the scheduling
policy used.  }

\vspace*{-2ex}
\label{fig:wrm}
\end{center}
\end{figure*}

As briefly discussed, each Worker process is able to use multiple computing
devices in a node. The devices are used cooperatively by dispatching fine-grain
tasks for execution in each CPU core or coprocessor.  Because multiple stage
instances may be active with a Worker, the tasks being computed may have been
created by different application stages.  In our implementation, fine-grain
tasks created by a stage are dispatched for execution with the Worker Resource
Manager (WRM) in each node (See Figure~\ref{fig:wrm} for details). The WRM
instantiates one computing thread for managing each CPU core or coprocessor.
Whenever idle, those computing threads inform the WRM, which selects one of the
tasks ready to execute with processing function implementation for the
targeting processor.  The scheduling policies used for selecting tasks are
described in Section~\ref{sec:opts}.

When all tasks dispatched for execution by a stage instance have finished, a
callback function is invoked to notify the WCT. WCT executes data staging and
writes the outputted region templates to appropriate global data storage.  A
message is then sent to the manager process with the information about the
completed stage. The Manager releases dependencies on that stage instance and,
as a consequence, other stage instances may be dispatched for execution.  The
stages of  region templates applications are developed in our framework as
implementations of a special region templates stage abstract class. This class
includes interfaces to insert, retrieve, and query region templates used by the
application stage. The runtime system also uses this interface to access region
templates associated to a given stage instance to (i)~stage/read global data
regions that are outputted/consumed and (ii)~delete region templates that will
not longer be used (input only and other regions after they are staged). Also,
the region templates stage class needs to provide mechanisms for
packing/unpacking itself to/from a data buffer when it is assigned to a Worker.

\subsubsection{Optimized Execution on Hybrid Systems}
\label{sec:opts}

This section details the optimizations implemented in our runtime system
targeting hybrid systems, which include smart task scheduling and strategies to
reduce impact of data transfers between CPUs and GPUs.

\paragraph{Performance Aware Task Scheduling (PATS)} As previously discussed,
the coarse-grain stages of the application may create several fine-grain task
or operations, which are dispatched for execution with the WRM. In real-world
data analysis applications, such as those that motivated the development of
region templates, we expect that several fine-grain tasks will be use to
compose the application. These tasks/operations tend to differ in terms of data
access pattern and computation intensity, thus they are also likely to attain
different speedups when executed on an accelerator.  In order to take advantage
of these performance variability, we have proposed and developed the PATS
scheduling~\cite{Teodoro-IPDPS2013,Teodoro-IPDPS2012} that is used with region
templates. PATS assigns tasks to a CPU core or a GPU based on the tasks
estimated acceleration on each device and on the processors load. Once tasks
are dispatched for execution with the WRM, they are either inserted in a list
of tasks pending or ready to execute. The pending tasks are those that do not
have all dependencies resolved, and the ready tasks are prepared for execution.
New tasks may be inserted in the ready tasks queue during the execution by the
application or because they had the dependencies resolved.  The PATS scheduler
maintains the list of tasks ready for execution sorted according to the
estimated speedup on a GPU, and the mapping of tasks to a CPU core or a GPU is
performed in a demand-driven basis as these devices become idle. If the idle
processor is a CPU core, the task with smallest speedup is selected for
computation, whereas the task with maximum speedup is chosen when a GPU is
available (See Figure~\ref{fig:wrm}).

The PATS scheduler only relies on tasks' speedup estimates that are accurate
enough to maintain correct order of tasks in the queue. Thus, it will not
suffer from inaccuracy in speedup estimates as long as it does high enough to
affect the order of tasks. There are several recent efforts on task scheduling
for hybrid
systems~\cite{mars,merge,qilin09luk,1616864,Diamos:2008:HEM:1383422.1383447.Harmony,6061070,ravi2010compiler,6152715,hpdc10george,6267858,Rossbach:2011:POS:2043556.2043579,Teodoro:2012:Cluster,augonnet:hal-00725477,cluster09george,hartley,HartleySC10,teodoro-2013-vldb,Hermann:2010:MMP:1885276.1885302,conf/ipps/GautierLMR13}.
Most of the previous works deal with task mapping for applications in which
operations attain similar speedups when executed on a GPU vs on a CPU. PATS, on
the other hand, exploits performance variability to better use heterogeneous
processors. Time-based schedulers (i.e., heterogeneous earliest finish time)
have been successfully used in heterogeneous system for a long time. These
class of schedulers, however, are very challenging to be employed in our
application. Several operations (ReconToNuclei, AreaThreshold, FillHolles,
PreWatershed, Watershed, etc) in our use case application have irregular
data-access and computation patterns with execution times that are
data-dependent.  Thus, execution times of those operations can not be
accurately estimated before execution. The speedup based scheduling used with
PATS has shown to be easier to apply, because we observed smaller variation in
speedups as consequence of input data variation.

\paragraph{Data Locality Conscious Task Assignment (DL)} The time spent
transferring data between CPU and GPU memories may have a great impact on the
overall execution time of a computation carried out on a GPU. Thus, it is
important for the scheduler to consider the location of the data used by a task
(CPU or GPU memory) in order to maximize performance. However, it should
minimize data movements without resulting in under utilization of a device,
because executing a task in a device the computes it inefficiently.  The data
used as input or generated as output of each task executed in our runtime
system are stored using region templates.  In addition, region templates
associated with a given task can be queried by the runtime system using the
task class API. Using this API, we have extended the basic scheduler to
incorporate data location awareness as an optional optimization. If it is
enabled, the scheduler searches the dependency graph of each task executed on a
GPU in order to identify tasks ready for execution that could reuse data
generated by the current task. If the speedups of the tasks are not known
(i.e., when First-Come, First-Served scheduler is used), DL forces the
scheduler to select one task that reuses data for execution when one exists. In
cases in which the speedup is known and PATS is used, it compares the
performance of tasks that reuse data with those that do not reuse. The task
with the best speedup in the queue ($S_q$) is compared to the task with the
best speedup that reuses data ($S_d$).  If ($S_d \ge S_q \times (1 -
TransferImpact$)), the operation that reuses data is chosen. TransferImpact
refers to the portion that the data transfer represents of the total task
execution time. This value is currently provided by the user. The reasoning of
this approach is that there may be operations that do not reuse data, but even
paying an extra startup cost make better use of a device. Thus, it may be more
appropriate to pay the data transfer costs than under utilizing a processor.
This optimization is also utilized in CPU-based executions to allow for
architecture-aware scheduling of tasks, which is an important optimization in
the context of current non-uniform memory architectures (NUMA) machines. Thus,
during the assignment of a new task for a CPU computation, the depending tasks
of the previously computed task are given a priority for scheduling on that
computing core. In our implementation, similarly to the case of a GPU, the
dependency graph of the current tasks is explored in order to find the
depending task that maximizes the amount of data reuse. This strategy asserts
good data reuse and reduces data movements.

\paragraph{Data Prefetching and Asynchronous Data Copy} Data prefetching and
Asynchronous Data Copy are other techniques employed by our runtime system to
mitigate the costs of data transfers. When DL is not able to avoid data
movements, the runtime system will try to perform these operations in
background to useful computation. In this approach, data used by tasks to be
executed on a GPU or outputted by previous tasks are transferred to/from GPU
memory concurrently to the computation of another task.  In our implementation,
the dataflow structure of the application is exploited once more to identify
the data regions that need to be transferred, and the region templates API is
used by the runtime system to perform the actual data transfers. The tasks
executed on a GPU are then pipelined through three phases: uploading,
processing, and downloading. In this way, data used by a task may be uploaded
simultaneously to the computation of a second task, and to the download of data
outputted by a third task.

\subsection{Region Templates Data Abstraction}
\label{sec:rt-data-abstraction}
The region template abstraction provides a generic container template for
common data types in microscopy image analysis, such as pixels, points, arrays
(e.g., images or 3D volumes), segmented and annotated objects and regions, that
are defined in a spatial and temporal domain.  With this abstraction, the
process of managing instances of the data types in a high performance computing
environment is pushed to the middleware layer, allowing for implementations of
different data storage and management capabilities, while providing a unified
interface to applications read and write these data types. A region template
instance represents a container for a region defined by a spatial and temporal
bounding box. A \emph{data region} object is a storage materialization of data
types and stores the data elements in the region contained by a region template
instance. A region template instance may have multiple data regions.
Application operations interact with data region objects to retrieve and store
data. That is, an application writes data outputs to data regions and reads
data inputs from data regions, rather than reading from or writing directly to
disk or directly receiving data from or sending data to another stage in the
workflow. 

Region templates and data regions can be related to other region templates and
data regions. Data regions corresponding to the same spatial area may contain
different data types and data products. For example, data regions can be
related to each other to express structures at different scales that occupy the
same space. Data regions can also be related to each other to express evolution
of structures and features over time. The spatial and temporal relationship
information may be used by the middleware layer to make decisions regarding
distribution and management of data regions in a high performance
computing environment. Data regions are identified by a {\em (namespace::key,
type, timestamp, version number)} tuple. This identifier intends to provide
temporal relationships among data regions related to the same spatial area.

The region template library provides mechanisms for defining region templates
and instantiating data regions with support for associative queries and direct
access to data elements. A single region template instance may contain multiple
data regions. A simplified version of the Region Template class definition is
presented in Figure~\ref{fig:RTClass}. Multiple data regions are stored into a
map of data regions. Data regions with same name are stored into a list and
they must differ by at least one of its identifiers: type, timestamp, and version
number. A given region template instance also contains a bounding box with
coordinates of the space it covers. As data regions are inserted into a region
template, the bounding box of the region templates is be updated such that it
remains the minimum bounding box to contain bounding boxes of data regions
inserted.

\begin{figure}[h!]
\begin{center}
\makebox[0.495\textwidth]{
 \subfigure[Region Templates Class.]{\label{fig:RTClass} 
\raisebox{2.2cm}{
\begin{lstlisting}
class RegionTemplate {
private:
  // Data regions: 1D/2D/3D regular
  // or irregular, and polygons
  std::map<std::string, 
     std::list<DataRegion*> > 
     templateRegions;

  // Region template name identifier
  std::string name;

  // Region template coordinates
  BoundingBox bb;

  // If false, data are automatically 
  // read during component creation
  bool lazyRead;
           ...
public:
  bool insertDataRegion(
    DataRegion *dataRegion);
  DataRegion *getDataRegion(
    std::string namespace, 
    int timestamp=0, int version=0, 
    int type);

  int getNumDataRegions();
  DataRegion *getDataRegion(int idx);
  void instantiateRegion();
          ...
};
\end{lstlisting}}
}}
\makebox[0.495\textwidth]{
 \subfigure[Data Region Abstract Class.]{\label{fig:DRClass} 
\raisebox{2.2cm}{
\begin{lstlisting}
class DataRegion {
private:
  // Type of each data element CHAR, UCHAR,...
  int elementType; 
  // Dense, sparse, (2D/3D),...
  int regionType;	 
  int version, timestamp;
  // Name and instance identifier  
  std::string name, id;
  // Resolution of the region    	
  int resolution;
  // BB surrounding domain and ROI  
  BoundingBox bb, ROI;  

  // Associates a given bounding box to an id, 
  // for data that may be split into pieces. 
  std::vector<std::pair<BoundingBox, 
      std::string> > bb2Id; 

  // Type of storage used to read: distributed 
  //shared memory and high performance disk
  int inputDataSourceType, outpuDataSourcetType;
	...
public:
  DataRegion();
  virtual ~DataRegion();
  virtual bool instantiateRegion();
  virtual bool write();
  virtual bool empty();
  	...
};
\end{lstlisting}}
}}
\vspace*{-1ex}
\caption{Simplified version of the Region Templates and Data Region
Abstractions. A Region Template data structure may store several data regions,
which are distinguished by their tuple identifier. The Data Region defines a
base class that is inherited by concrete implementations of different data
region types. Abstract methods in the Data Region class include those
operations that are type specific and need to the implemented to create a new
data region type. The system currently includes implementations of the following
data region types: 1D, 2D, or 3D dense or sparse regions, polygons. These
implementations also include node local implementations targeting shared memory
hybrid machines, equipped with CPUs and accelerators.} 
\vspace*{-2ex}
\label{fig:code_dpartition} 
\end{center} 
\end{figure}

The instantiation of the data regions stored in a region template used by an
application stage instance may or may not be performed when that instance is
received by the the Worker. If it is lazy, the actual data stored in each data
region is only read when it is first accessed. Therefore, the data regions
stored in a region template instance will only store the metadata describing
the data regions, which are necessary and sufficient to retrieve the actual
data from a global data storage implementation. In the master component of a
region templates application, for instance, the lazy mechanism is used by
default, because we do not want to read the data at that level.

At the Worker level, our default strategy, as described in
Section~\ref{sec:runtime}, is not to use lazy strategy and data used by a
component instance is read by the Worker before the component execution. The
user is able to subscribed this default rule, and request for data reading to
be lazy. The latter approach is useful, for instance, if data accessed by a
component does not fit entirely in memory or if the bounding box to be read is
only known at runtime. In both cases, however, the data read during a data
region instantiation are that within its Region of Interest (ROI) bounding box.
This approach allows for a region template to store metadata of large data
regions, which are split for parallel computation by simply modifying the ROI.

Access to a data region in a region template instance is performed through
get/set operations and the data region tuple identifier. The different data
regions types implemented must inherit from the DataRegion abstract class
(presented in Figure~\ref{fig:DRClass}).  In addition to the tuple identifier,
the DataRegion class also includes attributes such as the type of elements
stored, type of the region, etc. The type of the region, for instance,
describes whether it stores 1D, 2D, or 3D dense or sparse regions, polygons,
etc. For each of these types, we have a different implementation of the DataRegion class, because the methods for instantiating/writing those different types
from/to the global storage differ from according to the data structure used.
New data region types may be added by implementing the DataRegion class.

Most of the data regions currently supported are implemented using
OpenCV~\cite{opencv_library} underneath. Such as
ITK~\cite{ITKSoftwareGuide,ITKSoftwareGuideThirdEdition}, OpenCV supports a
number of data types that include Points, Matrices, Sparse Matrices, etc, which
are commonly used by image analysis applications. An additional feature of
OpenCV that motivated its use is the support for GPUs, which includes some of the
data structures and processing methods. 

The currently supported data region types: 1D, 2D, or 3D dense or sparse
regions, polygons, have implementations for local storage in a shared memory
machine targeting hybrid systems. As such, the region templates data structures
have both CPU and GPU based counterpart implementations, allowing for region
templates instances and their data structures to resided into local the CPU or
CPU memory of a node. It includes capabilities for moving the data structures
between the CPU and GPU memories through data uploads and downloads interfaces.
Data transfers in each direction may be carried synchronously or
asynchronously. The region template interface includes blocking and
non-blocking query operation to verify whether a transfer has finished. As
discussed in Section~\ref{sec:opts}, the runtime system takes advantage of
asynchronous transfer mechanisms to overlap data transfers with useful
computation.

Data regions exchanged among application components, which may run on different
nodes of a distributed memory computed, are called global data regions. The
current implementation of the system provides different implementations of
\emph{data storage} for global regions. The currently supported implementations
are detailed in Section~\ref{sec:storage-impl}, and include high performance
disk and distributed shared memory storage.

\subsection{An Example Region Templates Application Workflow} 
\label{sec:compose}
An application in our analytics framework consists of a manager component and
the application stages that are executed with Workers. The manager (master)
component specifies the analysis workflow and defines the region templates for
the application.  The application developer needs to provide a library of data
analysis operations and to implement the portions of the manager that specify the
analysis pipeline and region templates used in each stage. In this process, the
manager component may need to partition each region encapsulated by the
region templates among workflow stage instances for parallel execution. Our
current implementation supports arbitrary partitions of the data regions.

Figure~\ref{fig:managerCode} presents a sketch of the manager code in our use
case application. It defines the libraries in which the application stages are
implemented. Further, in our specific application case, the manager component
reads a list of image tile file names from an input folder and builds a single
region template instance ("Patient") from all tiles. In this example, a data
region ("RGB") is created and each image tile is inserted in the data region
with appropriate bounding box of the subdomain it refers to. As presented in
Section~\ref{sec:rt-data-abstraction}, a data region may be partitioned into
several chunks of data with associated bounding box (See $bb2Id$ structure in
Figure~\ref{fig:DRClass}).

\begin{figure*}[htb!]
\begin{center}
\includegraphics[width=1\textwidth]{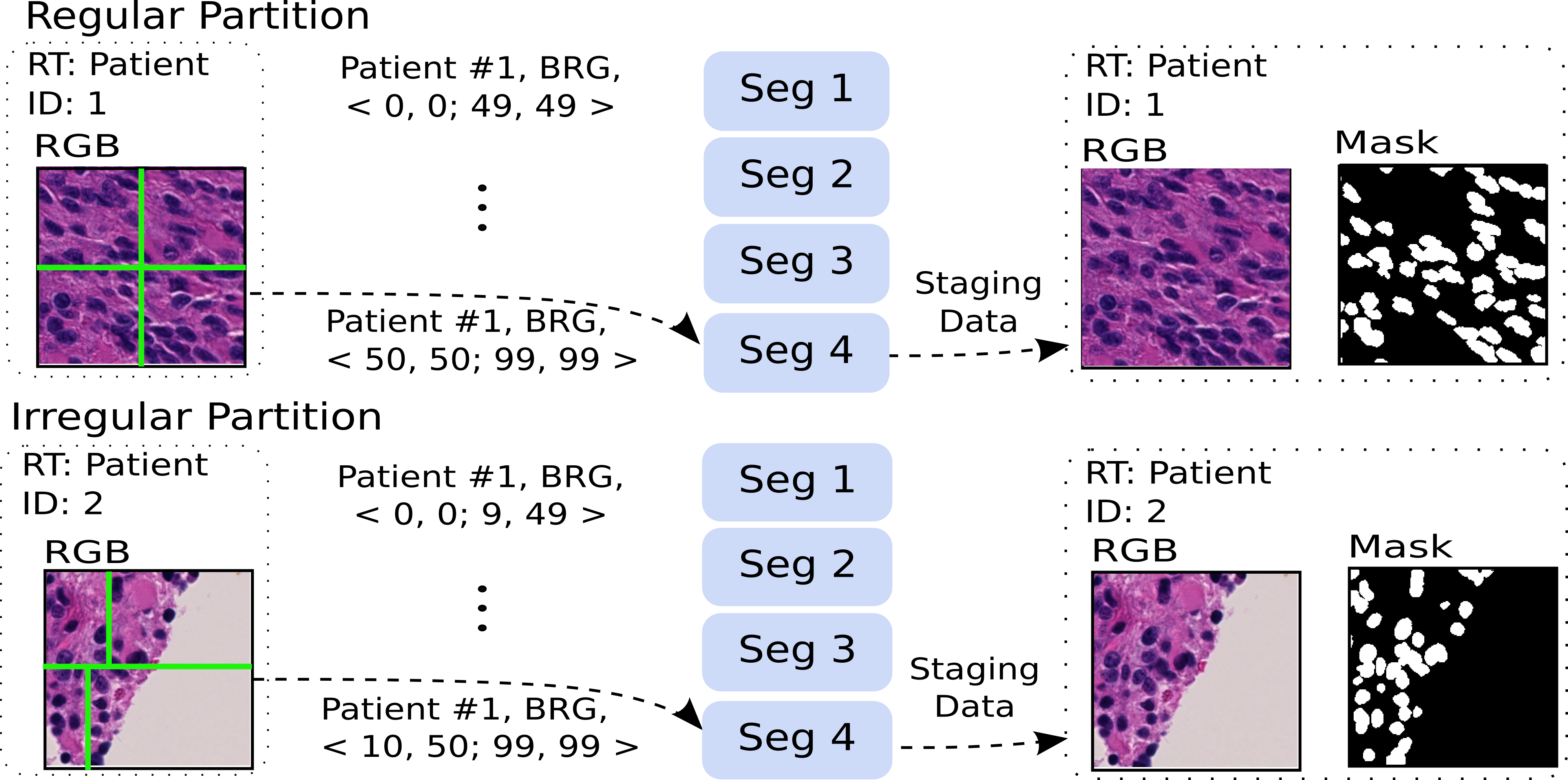}
\vspace*{-1ex}
\caption{Two possible data partitions and instantiations of
that stage, using regular (blocks of 50$\times$50 pixels) and irregular blocks for
better balancing of computational load. Irregular partitions would
be useful, e.g., in computations of unstructured grids.}
\vspace*{-2ex}
\label{fig:dpartition}
\end{center}
\end{figure*}

After the data region is created, our manager code partitions the data domain
of the data region used as input. The user may create arbitrary partitions that
are more appropriate for her application. As an example, two partition
approaches could be used with our application (See Figure~\ref{fig:dpartition})
-- the code segment that creates partitions is not shown. In the regular case,
input data region are partitioned into 50$\times$50 tiles. In the irregular
data partition, a customized partition is implemented based on some metric
provided, for instance, object or tissue density.

Further, the manager code creates a copy of each application stage per
partition, and sets data regions used in each of them. The data region
information includes the appropriate ROI of that instance, type of data region
(input, output, input and output), and global storage used. During this
process, the dependencies among stage must also be provided by the application
developer.  Finally, the stage instances are dispatched for execution with the
runtime system, which waits until the application completes.

\begin{figure*}[htb!]
\begin{center}
\mbox{
 \subfigure[Manager component of the application.]{\label{fig:managerCode} 
\raisebox{2.2cm}{
\begin{lstlisting}
int main (int argc, char **argv){
	...
  RegionTemplate *rt;
  // Libraries that implement components
  SysEnv sysEnv(argc, argv, "libcomponentsrt.so");
 
  // Create a region templates instance. In this example, a single 
  // input image is stored into multiple files. Each file is inserted 
  // into the same data region with appropriate bounding box.
  rt = RTFromFiles(inputFolderPath);

  // Create partition of the RT for parallel execution
  vector<BoundingBox *> partitions = createPartition(rt);

  // Instantiate application dependency graph
  for(int i = 0; i < partition.size(); i++){
    // Create an instance of Segmentation stage
    Segmentation *seg = new Segmentation();
    // Insert region template to be used, with it ROI:partitions[i]
    seg->addRegionTemplateInstance(rt, "RGB",  partitions[i], INPUT_OUTPUT, DISK);
    seg->addRegionTemplateInstance(rt, "Mask", partitions[i], OUTPUT, DMS);

    // Cretate instance of Feature Computation stage
    FeatureExtraction *fe =  new FeatureExtraction();
    fe->addRegionTemplateInstance(rt, "RGB",  partitions[i], INPUT, DMS);
    fe->addRegionTemplateInstance(rt, "Mask", partitions[i], INPUT, DMS);

    // Set stage dependencies 
    fe->addDependency(seg);
    // Dispatch stages for execution
    sysEnv.executeComponent(seg);
    sysEnv.executeComponent(fe);
  }
  // Start Executing
  sysEnv.startupExecution();
  // Wait until application is done 
  sysEnv.finalizeSystem(); 
	...
}
\end{lstlisting}}
}}
\mbox{
 \subfigure[Simplified code for the segmentation stage.]{\label{fig:segCode} 
\raisebox{2.2cm}{
\begin{lstlisting}
int Segmentation::run()
{
  // Get region template (RT) handler
  RegionTemplate *inRT = getRegionTemplate("Patient");

  // Get data region inside that RT
  DenseDataRegion2D *rgb = 
	dynamic_cast<...>(inRt->getDataRegion("RGB"));

  // Change version of the outputted region 
  rgb->setVersion(rgb->getVersion()+1);

  // Create output data region
  DenseDataRegion2D *mask = 
	new DenseDataRegion2D("Mask", DMS);
  mask->setBb(rgb->getBb());
  inRT->insertDataRegion(mask);

  // Create processing task
  Segmentation(rgb, mask);
}

\end{lstlisting}}
}}
\vspace*{-1ex}
\caption{Simplified code of (a)~manager component and (b)~segmentation
stage in our example application. The manager component is responsible for
setting up the runtime system, initializing the metadata of the region
templates and data regions used, partitioning the data regions for parallel
execution, and creating the application graph with appropriate dependencies
among stage instances. The segmentation code represent the interactions of an
application stage with region templates to retrieve data, and transform it. As
discussed, the data staging of output data regions is handled by the runtime
system.} 
\vspace*{-2ex}
\label{fig:code_dpartition} 
\end{center} 
\end{figure*}

Figure~\ref{fig:segCode} presents the code for the segmentation stage used in
our example analysis application. A region template, called "Patient", defines
a container for data regions "RGB" and "Mask". The data region "RGB" is
specified as input and output, meaning it is read from a data region storage
and written to a storage at the end of the stage.  The data region "Mask" is
specified as output, since it is produced by the segmentation stage.  When a
copy of the segmentation stage is executed on a node, it interacts with the
runtime system through the region template. It requests the data region "RGB",
creates the data region "Mask", associates "Mask" with data distributed memory
storage implementation "DMS", and pushes it to the region template instance. We
should note that the code in the figure is simplified for space reasons. As
described earlier, the runtime system have access to the data regions and
region templates used by a stage, and is able to instantiate the data region
used and makes it available to the segmentation stage (see
Figure~\ref{fig:execModelRuntime}). 

In this example, a copy of each stage is executed per partition.  Assume the
bounding box of the entire region is ($<0, 0; 99, 99>$).  The bounding box of
partition 4 of the data region "RGB" is defined as $<50, 50; 99, 99>$, a copy
of the segmentation stage, e.g., "Seg 4" in Figure~\ref{fig:code_dpartition} is
created to compute that region.  Preserving the bounding box of the original
region template is important to allow for the application to identify the
location of the data regions in the original data domain. This information is
useful, for instance, for computations involving overlapping borders.  The case
of ghost cells, for instance, may be handled in a region templates application
by (i)~creating ROIs that include the ghost cells during the process of reading
data and (ii)~shrink the ROIs of data regions to remove the ghost cells before
they are staged.

As noted earlier, data regions can be associated with different data storage
implementations. In the code segment in Figure~\ref{fig:segCode}, the data
region "RGB" is read from the high performance disk ("DISK") and written to
"DMS" storage implementation. If at least one of the tuple identifiers of the
region is not modified, two copies of the same data region would exist in
different global storage implementations after segmentation.  In this case,
unless otherwise specified by the user, the system will use the lattest staged
region in the remaining references to that data region.  The case in which
multiple stage instances reads and stages overlapping areas of a data region to
the same data storage implementation may also lead to synchronization problems.
The global data storage implementation always keeps the last staged version of
overlapping data regions. Therefore, the application developer should set
dependencies among stages correctly to avoid such issues.

\section{Global Storage Implementations for Data Regions}
\label{sec:storage-impl}
We have developed implementations of data regions that represent data
structures used in the object segmentation and feature computation operations.
Input to an object segmentation operation is an image tile, output from the
segmentation operation is a mask array. Input to a feature computation
operation is a pair of (image tile, mask array), while the output of a feature
computation operation is a set of feature vectors.  Each feature vector is
associated with a distinct segmented object. The implementations include local
and global storage. The local storage, detailed in
Section~\ref{sec:rt-data-abstraction}, refers to region templates and data regions local to a node
stored in CPU and/or GPU memories. The global storage is accessible by any node
in the system and stores global regions that are used to exchange data among
the analysis stages of an application. The storage implementations for global
data regions are presented in Sections~\ref{sec:dms} and~\ref{sec:ds}.

\subsection{Distributed Memory Storage} \label{sec:dms}
\begin{figure}[h!]
\begin{center}
   \includegraphics[width=1\textwidth]{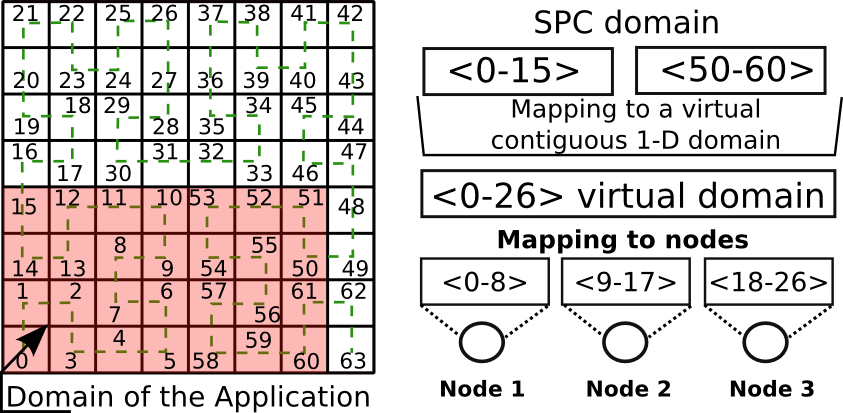}
   \vspace*{-3ex}
   \caption{Space-filling curve based distributed hash table (DHT). The application
   2-dimensional domain (highlighted) is first mapped in a contiguous virtual domain, and is
   further divided to be stored in multiple storage nodes.} 
   \vspace*{-2ex} 
   \label{fig:dht} 
\end{center} 
\end{figure}

The distributed memory storage (DMS) implementation is built on top of
DataSpaces~\cite{DocanPK10}. DataSpaces is an abstraction that implements a
semantically specialized virtual shared space for a group of processes. It
allows for data stored into this space to be indexed and queried using an
interface that provides dynamic and asynchronous interactions. The data to be
retrieved/stored is described through key-value pairs that define a bounding
box within the application address space.  DataSpaces is implemented as a
collection of data storage servers running on a set of nodes. An important
component of DataSpaces is its distributed hash table (DHT) presented in
Figure~\ref{fig:dht}, which has been designed as an indexing structure for fast
lookup and propagation of metadata describing the data stored in the space.  In
multi-dimensional geometric domains, DataSpaces employs Hilbert space-filling
curve (SFC)~\cite{Bially-1976-PhdThesis} to map $n$-dimensional points to an
1-dimensional domain for storage in the DHT.  The resulting 1-dimensional SPC
domain for an application data may not be contiguous and, as such, it mapped
into a virtual domain before data is distributed among the storage nodes.

The systems is built using a layered architecture with communication, data
storage, and data lookup layers. The communication layer provides a set of
communication primitives, which are implemented on top of advance network
technologies used in high-performance systems. It includes Remote Direct Memory
Access (RDMA) and one-sided communication. This layer also defines protocols
for communication between application and storage system, and between components
of the storage system. The data storage layer is responsible for storing data received
from the application and managing memory allocation and buffering. Data
received from the application is stored in-memory. This layer complements the
DHT that maintains metadata describing location of the data.  Finally, the data
lookup layer provides the mapping of application domain geometric descriptors
in keys of the DHT (See Figure~\ref{fig:dht}), as well as it is coordinates the
routing of queries to nodes containing the data.

Our implementation provides a specialized factory object, which can stage data
regions from a region template instance to DataSpaces and retrieve data regions
to create local instances of the data regions on a node. In the process of
staging a region template instance (and the data regions it stores) to
DataSpaces, a DataSpaces insertion request (formed using the data region tuple
identifier and the data region bounding box) is created for each data region in
the region template instance.  Then, the data into the data regions are packed
according to the application domain description and the system dispatches
asynchronous insertion requests to the space.  Similarly, in the read
operation, the read requests are created from an identifier and bounding box
describing the portion of the data domain to be retrieved and DataSpaces is
queried in background.  After data are returned, the system instantiates the
application local CPU-based data regions and associates them to the region
template instance being created. As discussed before, the computation stages
are dispatched for execution only after the data is retrieved and the data
regions are locally available. This allows for the data movement necessary for
the execution of a stage instance to be performed in background to other stage
instances execution.

\subsection{High Performance Disk Storage} \label{sec:ds}
We have developed an implementation for disk storage based on a stream-I/O
approach, drawing from filter-stream
networks~\cite{ARPACI-99,PlaleS03,KumarSMVDRKGHKS09} and data
staging~\cite{DocanPK10,AbbasiWEKSZ10}.  The storage implementation assumes
that data will be represented and staged to disk in application-defined chunks.
In our current implementation of image analysis pipelines, the processing of a
set of images in the segmentation and feature computation stages is carried out
in image tiles. Output from the segmentation stage is a set of mask arrays,
each of which corresponds to an image tile. A data region with \emph{global}
scope is used to store these mask arrays on a distributed memory machine. The
I/O component provides the data storage abstraction associated with the data
region. When a computation node finishes processing data associated with an
image tile, it writes the mask array to the data region.  The data region
passes the output as a data chunk to the data storage abstraction layer.  This
approach allows us to leverage different I/O configurations and sub-systems.
In the current implementation, in addition to POSIX I/O in which each I/O node
can write out its buffers independent of other I/O nodes, we have integrated
ADIOS~\cite{LofsteadKSPJ08} for data output. ADIOS is shown to be efficient,
portable, and scalable on supercomputing platforms and for a range of
applications. 

In our implementation, I/O operations can be co-located with computation
operations, and we have extended ADIOS to support the case in which a set of
CPU cores can be designated as I/O nodes.  In the co-located case, we have also
developed support to allow for processors to be partitioned into I/O clusters,
whereas in the original ADIOS implementation only a single I/O cluster is used
for all processors. All the processors in the same I/O cluster use the same
ADIOS group and may need to synchronize. Each I/O cluster can carry out I/O
operations independent of other I/O clusters; there is no synchronization
across I/O clusters. This clustering is aimed at reducing the synchronization
overheads. In the separate I/O-computation configuration, the I/O nodes are
coupled to the computation nodes via logical streams. The I/O nodes are further
partitioned into groups of $k$ I/O nodes -- all the I/O nodes could be in the
same group ($k = N$, where $N$ is the number of I/O nodes), or each group could
consist of a single I/O node ($k = 1$). In this setting, when a computation
node is ready to output a data chunk, it  writes the mask array to its output
stream. The stream write operation invokes a scheduler which determines to
which I/O node the data buffer should be sent, and sends the buffer to the
respective I/O node. When an I/O node receives a buffer from its input stream,
it puts the buffer into a queue. When the number of buffers in the queue in an
I/O node reaches a predefined value, all the I/O nodes in the same group go
into a write session and write the buffers out to disk.  This implementation
facilitates flexibility. The I/O nodes can be placed on different physical
processors in the system. For example, if a system had separate machines for
I/O purposes, the I/O nodes could be placed on those machines. Moreover, the
separation of I/O nodes and computation nodes reduces the impact on computation
nodes of synchronizations because of I/O operations and allows a scheduling
algorithm to redistribute data across the I/O nodes for I/O balance. We have
implemented round-robin and random distribution algorithms. In summary, this
module of our system extends ADIOS to support (i)~separated I/O cores and
(ii)~configurable I/O cluster/group sizes, since original ADIOS have a single
cluster that contains all processes.

\section{Experimental Results} \label{sec:res}
\begin{figure}[htb!]
\begin{center}
\includegraphics[width=0.85\textwidth]{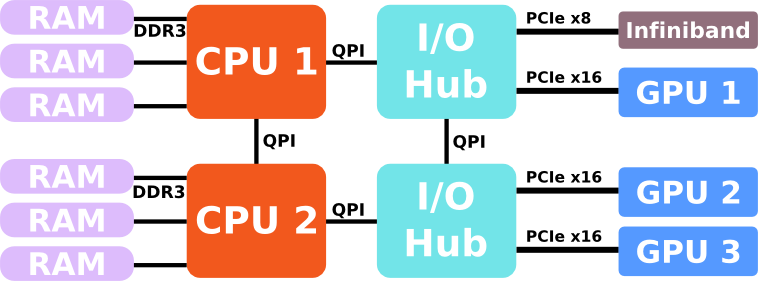}
\vspace*{-1ex}
\caption{Keeneland computing node architecture.}
\vspace*{-2ex}
\label{fig:keeneland}
\end{center}
\end{figure}

We have evaluated the region template framework using the Keeneland distributed
memory hybrid cluster~\cite{10.1109/MCSE.2011.83}. Keeneland is a National
Science Foundation Track2D Experimental System and has 120 nodes in the current
configuration. Each computation node is equipped with a dual socket Intel X5660
2.8 Ghz Westmere processor, 3 NVIDIA Tesla M2090 (Fermi) GPUs, and 24GB of DDR3
RAM (Figure~\ref{fig:keeneland}). The nodes are connected to each other through
a QDR Infiniband switch. The image datasets used in the evaluation were
obtained from brain tumor studies~\cite{ieee-insilico}. Each image was
partitioned into tiles of 4K$\times$4K pixels, and the background only tiles
were removed from the tile set. The codes were compiled using ``gcc 4.4.6'',
``-O3'' optimization flag, OpenCV 2.3.1, and NVIDIA CUDA SDK 4.0. The
experiments were repeated 5 times.  The standard deviation in performance
results was not observed to be higher than 3\%. The input tiles were stored in
the Lustre file system attached cluster. 

\subsection{Example Application}
The example application implements a segmentation stage and a feature
computation stage, as shown in Figure~\ref{fig:app}
(Section~\ref{sec:motivating}). The segmentation stage receives a RGB image as
an input and produces a mask identifying segmented nuclei. The feature
computation stage computes a set of shape and texture features for each
segmented nucleus.  Each stage is composed of a series of functions with the
CPU and GPU implementations.  For functions such as Morphological Open, Color
deconvolution, Canny, and Gradient, we have used the implementations in
OpenCV~\cite{opencv_library}.  The GPU based version of Watershed is based on
the implementation by K\"{o}rbes et.
al.~\cite{Korbes:2011:AWP:2023043.2023072}.  The functions "ReconToNuclei",
"FillHoles", and "Pre-Watershed" were implemented using the irregular wavefront
propagation pattern (IWPP) optimized for GPU execution by our
group~\cite{Teodoro:2013:Parco}. The IWPP algorithms perform computation only
on a subset of elements (active elements) from the input data domain, since
other elements do not contribute to output. Our GPU implementation of the IWPP
algorithms uses a hierarchical and scalable queue to store and manage active
elements for high performance execution. We refer the reader to the earlier
work for the implementation details~\cite{Teodoro:2013:Parco}. The connected
component labeling function (BWLabel) is implemented by us based on the
union-find pattern described in~\cite{Oliveira10}. This pattern creates a
forest in which each pixel is a tree, and iteratively merges adjacent trees
with same mask pixel value. After the merging phase, labels can be extracted by
flattening the trees.  

The feature computation stage has relatively regular computation patterns and
achieves better GPU acceleration overall than the segmentation stage does.  For
feature computations on objects, we restructured the computation into a set of
smaller regions of interest: a set of minimum bounding boxes, each of which
containing a nucleus. Since each bounding box can be processed independently, a
large set of fine-grained tasks will be created in this strategy. By
restructuring the computations in this way, we avoid unnecessary computation in
areas that are not of interest (i.e., that do not have objects) and create a
more compact representation of the data. 

The parallel computation of object features using a GPU is organized in 2
steps. First we assign one GPU block of threads to each bounding box, and those
threads in a block collectively compute the intermediate results, i.e.,
histograms and co-occurrence matrices of the corresponding nuclei. By
structuring the computation in this way, we take advantage of a GPU’s ability
to dynamically assign thread blocks to GPU multiprocessors, in order to reduce
load imbalance that may arise due to differences in sizes and, consequently,
computation costs of different nuclei. In the second step, the nuclear feature
values are calculated from intermediate results, which are now fixed sized per
nucleus, and one GPU thread is executed per nucleus. A single step computation
of intermediate results and features is in contrast less efficient because:
(i)~the number of threads needed for intermediate results computation is much
higher than the number of features to compute, resulting in idle threads; and
(ii)~the computation of different features by threads in same block creates
divergent branches. Both inefficiencies are resolved with the two-step
approach.

\subsection{Single Node Performance} \label{sec:singe-node}
These experiments intend to quantify the overhead of using the region template
abstraction and the scalability of the example application on a single node of
the cluster. The timings reported in our results are the end-to-end runs and,
therefore, include the I/O costs of reading the input images from the file
system. 

\begin{figure}[htb!]
\begin{center}
\includegraphics[width=0.85\textwidth]{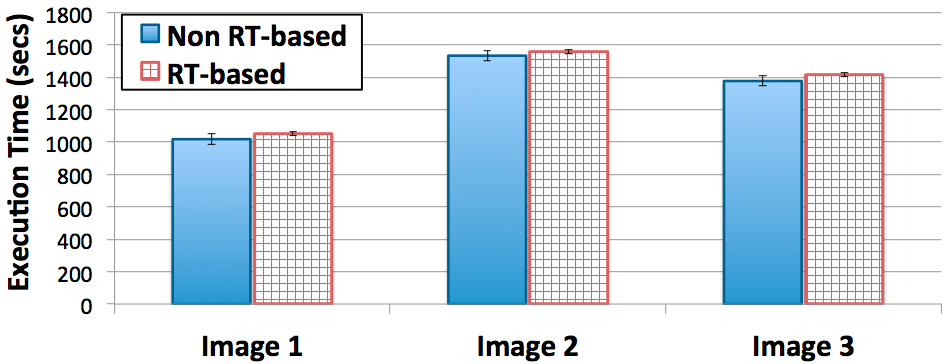}
\vspace*{-1ex}
\caption{Comparison of the Non-RT-based and RT-based versions on a single CPU core.}
\vspace*{-2ex}
\label{fig:overhead}
\end{center}
\end{figure}

We have developed a sequential single core version of the application (referred
to as {\em non RT-based}) and compared its execution times with those of the
single core version of the application using region templates (referred to as
{\em RT-based}). Both implementations were executed using 3 randomly selected
images: Image~1, Image~2, and Image~3, containing 108, 154, and 117
4K$\times$4K tiles, respectively. As shown in Figure~\ref{fig:overhead}, the
non RT-based version is only 1.03$\times$ and 1.02$\times$ faster,
respectively, in the best and average cases. This shows the overhead of the
region template abstraction is very small. 

The speedup values for the RT-based version on multi-core CPU and multi-GPU
executions, as compared to the single CPU core version, are presented in
Table~\ref{tab:single-node-table}. We first evaluate the impact of using the
Data Locality Conscious Task Assignment (DL) to the scalability. When this
optimization (described in Section~\ref{sec:opts}) is enabled, the system will
preferably assign to a given CPU core tasks that are depending on the
previously computed task on the same core, because those tasks will reuse data
and avoid data movement in the memory hierarchy. As presented, the multi-core
version of the application without DL achieved executions are slightly
sub-linear, i.e., 10.1$\times$ for the 12 cores configuration. This is a
consequence of the application's high memory bandwidth demand, which limits the
multi-core scalability as the threads compete for the shared memory subsystems,
and the higher costs for reading data as the number of computing cores
increase. Additionally, the use of DL to reduce the amount of that transfers
resulted into an application speedup of 10.9$\times$ on top of the single core
runs (with 12 cores used), and an improvement of about 1.08$\times$ as compared
to the multi-core counterpart without this optimization.

\begin{table}[h!]
\begin{center}
\begin{small}
\subtable[Parallel CPU-based executions with/without data locality conscious task assignment (DL).]{\label{tab:single-node-cpu}
\begin{tabular}{ c|c|c|c|c|c|c|c }
  \hline 
  \textbf{\# of CPU cores} 	& 1 	& 2	& 4	& 6	& 8	& 10	& 12	\\ \hline \hline
  \textbf{Speedup} 		& 1 	& 1.9	& 3.8	& 5.7	& 7.5	& 9.2	& 10.1	\\  \hline
  \textbf{Speedup - DL} 	& 1 	& 1.9	& 3.9	& 5.8	& 7.9	& 9.8	& 10.9	\\  \hline
\end{tabular}
}
\subtable[Multi-GPU scalability]{\label{tab:single-node-gpu}
\begin{tabular}{ c|c|c|c }
  \hline
  \textbf{\# of GPUs} 		& 1 	& 2	& 3	\\ \hline \hline
  \textbf{Speedup} 		& 7.9 	& 15.3	& 22.2	\\  \hline
\end{tabular}
}
\vspace*{-1ex}
\caption{Multi-core and multi-GPU scalability of the example application.}
\vspace*{-2ex}
\label{tab:single-node-table}
\end{small}
\end{center}
\end{table}

The speedups achieved by the multi-GPU executions of the RT-based version are
also presented in Table~\ref{tab:single-node-table}. Speedups of 1.94$\times$
and 2.82$\times$ on two and three GPUs, respectively are achieved with respect
to the single GPU version. The good scalability was obtained through a careful,
architecture-aware placement of threads managing GPUs. In this placement, the
GPU manager thread for a GPU is bound to the CPU core that is closest to the
GPU in terms of number of links to be traversed when accessing that GPU.
Without this placement, the speedup on 3~GPUs was only 2.27$\times$. 

\subsection{Disk Storage: Region Template High Performance Staging to Disk} \label{sec:diskEval}

These experiments evaluate the disk storage implementation for high speed
staging to disk. This module of our system is built on top of ADIOS, which is
extended in the Region Templates implementation to include support for
(i)~separated I/O cores and (ii)~configurable I/O group/cluster sizes. 

The evaluation was carried out on a large scale cluster, called Titan. Titan is
a US Department of Energy System with 18688 nodes, each with a 16-core AMD
Opteron 6274 processor, and 32 GB of memory.  Disk storage is provided through
a shared Lustre file system. While the other experiments were carried out using
Keeneland, this set of experiments used Titan because ADIOS is installed on
Titan for production use, Titan is attached to a more scalable storage system,
and we were able to use more CPU cores on Titan than on Keeneland.  

We used two I/O configurations. In the first configuration, called {\em
co-located I/O}, each CPU core is also an I/O node and performs both
computation and I/O operations. In the second configuration, referred to here
as {\em separated I/O}, each CPU core is designated as either a compute core or
an I/O core; the compute cores send the output data to the I/O cores for file
system writes. The co-located I/O configuration maximizes the number of cores
performing I/O, but introduces synchronization during I/O operations. The
separated I/O configuration insulates the compute cores from the
synchronization overhead, but introduces additional communication costs. We
investigated the effects of using each configuration along with different
transport mechanisms. Three transport mechanisms were tested: POSIX, where the
data are written to the file system independently via standard POSIX calls; MPI
LUSTRE, where ADIOS is aware of the Lustre parameters for the file target; and
MPI AMR, which is similar to MPI LUSTRE, but a staging layer is introduced to
increase data size per I/O request.  For each transport mechanism, we
partitioned the set of cores participating in I/O into MPI groups of size 1,
15, or the full I/O core size (ALL) to balance between synchronization impact
and transport mechanism requirements.  For the Separated I/O configuration, we
dedicated 60, 512, or 1536 cores for the I/O tasks. Each parameter combination
was run in triplicate using 2048 cores with 10240 4K$\times$4K input image tiles.

\begin{figure}[h!] 
\begin{center}
	\includegraphics[width=0.99\textwidth]{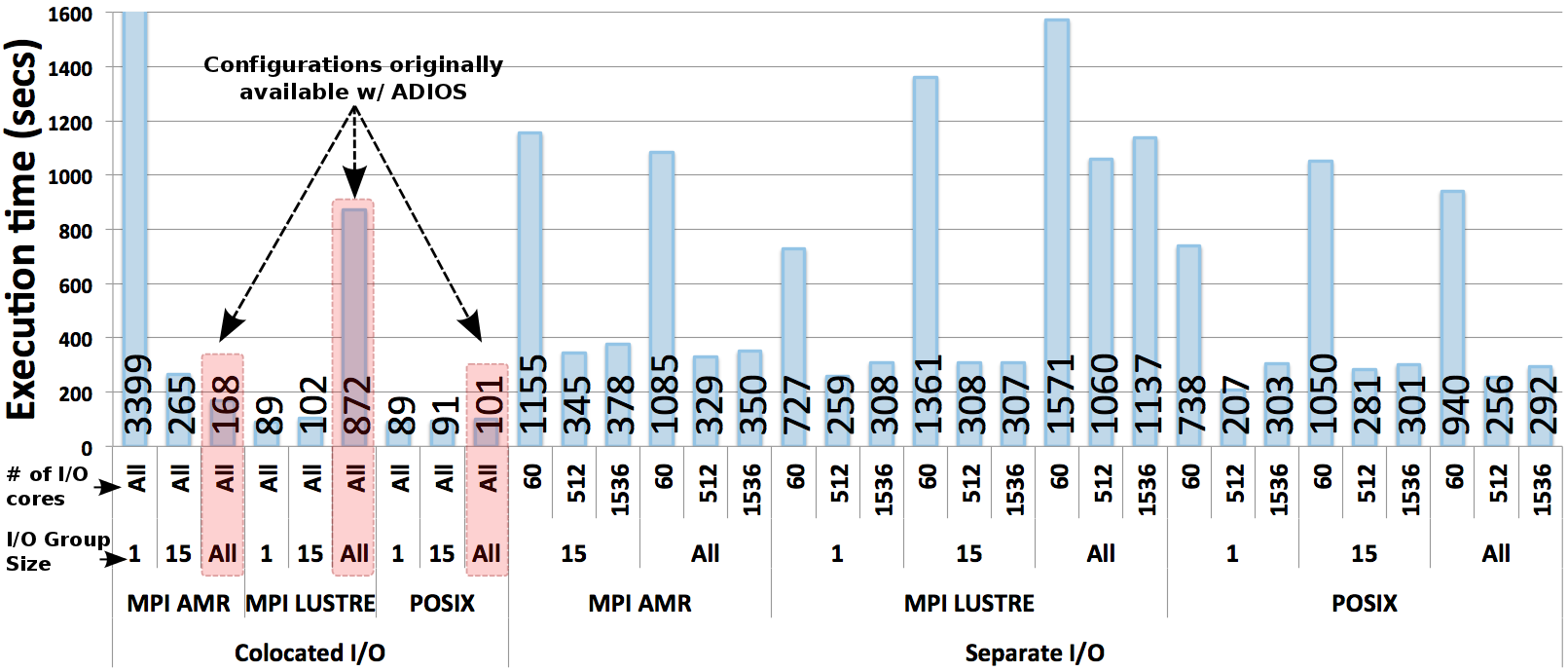}
	\vspace*{-1ex} 
\caption{Evaluation of high performance disk staging implementation on Titan.
Two configurations are tested for data staging: co-located I/O in which each
computing core also performs I/O operations, and separated I/O in which each
core is either a compute core or an I/O core.  Different transport mechanisms
may be used for each configuration: MPI AMR, MPI LUSTRE or POSIX.  Further,
Region Templates also implements the ability of using a configurable number of
cores in I/O groups (ADIOS originally only supports the configuration with All
cores in a single I/O group). The I/O group sizes evaluated are presented on
top of the mechanism chosen. Finally, the number of cores used as I/O cores is
presented in the labels closest to the X-axis. The configurations available
with the original implementation of ADIOS are highlighted. The remaining of the
configurations are a result of the extension implemented with Region Templates:
separate I/O and flexible I/O group sizes.  As presented, co-located I/O leads
to better performance in all configuration. Further, The MPI LUSTRE and POSIX
with co-located I/O attained to the best performance using
small I/O group sizes. As compared to the best original ADIOS configurations
(co-located and group size equals to All), the extensions with the Region
Templates resulted into a speedup of 1.13$\times$ in the application runtime.}
\vspace*{-2ex}
\label{fig:adios}
\end{center}
\end{figure}

Figure~\ref{fig:adios} shows that the co-located I/O configuration performs
better than the separated I/O configuration for all experiments. The
experiments with the co-located I/O configuration experienced decreased
performance when group size was increased for POSIX and MPI LUSTRE, showing
that support for smaller I/O groups implemented as part of the Region Templates
framework consistently improved the performance of the application -- the
default setup (originally implemented in ADIOS) supports only the configuration
with All processors in an I/O group. For MPI AMR and co-located I/O, we have
observed an opposite trend, as smaller groups would perform very poorly due to
overheads introduced without staging benefits.

For MPI AMR in the separated I/O configuration, we excluded group size of 1 as
this configuration produced extremely poor performance.  For most of the
separate I/O results, allocating 512 cores to I/O resulted in better
performance than 60 or 1536 I/O cores, potentially due to better balancing
between data generation rate at the compute cores, data transmission rate
between cores, and data consumption rates at the I/O cores. The configurations
with 60 cores for I/O resulted in lower performance, because of communication
contention when sending data to the I/O cores.  The MPI LUSTRE transport showed
a significant decrease in performance with the ALL group size, since it incurs
significant synchronization costs. 

Even though the separated I/O attained lower performance than co-located I/O,
we expect that it will improve the performance of the co-located I/O in other
scenarios. For instance, if we were able to run the separated I/O nodes into
the storage nodes, it would reduce the communication traffic. In that case, the
application would benefit from the asynchronous I/O supported by separated I/O,
because it caches data from I/O operations in memory, and performs the proper
write operations to storage in background to the application execution.

The region template abstraction allows us to choose different I/O
implementations and co-locate or separate the I/O nodes to achieve good
performance. The use of small I/O group sizes supported by Region Templates
resulted into an speedup of 1.13$\times$ on the application execution time as
compared to the best configuration originally available in ADIOS (Co-located
I/O, POSIX, and I/O group size All). We intend to examine in a future work
methods for automating the choice of the I/O configuration through the
integration with parameter auto-tuning
systems~\cite{DBLP:conf/sc/TapusCH02,DBLP:conf/ipps/YiSYVQ07,Nishtala2011576}.

\subsection{Performance of Distributed Memory Storage Implementation}
This section evaluates the performance of the distributed memory storage (DMS),
and compares it to the high performance disk storage (DISK) to exchange data
between the segmentation and feature computation stages. In the DISK storage,
I/O nodes and compute nodes were co-located with a group size of 1 (i.e., each
compute node performs I/O operations independently of the other nodes) and the
POSIX I/O substrate was used for read and write operations. This configuration
resulted in the disk storage best performance, as detailed in
Section~\ref{sec:diskEval}.   

In these experiments, the segmentation stage receives a region template with a
data region named ``RGB'' as input, and creates an additional data region named
``Mask" that identifies segmented nuclei. The DISK version of the application
reads the ``RGB'' from the file system and stages the ``Mask'' data region to
the file system at the end of the segmentation stage.  Both data regions are
then read from the file system during the feature computation stage. The DMS
version also reads the input ``RGB'' data region from the file system in the
segmentation stage. At the end of the segmentation stage, however, the ``RGB''
data region is staged to DataSpaces along with the ``Mask'' (the ``RGB'' data
region is marked as INPUT\_OUTPUT as presented in Figure~\ref{fig:segCode}).
Therefore, in the DMS version, the feature computation stage reads both data
regions directly from DMS.  The results presented in this section are from
strong scaling experiments, in which the size of the data used as input and the
number of nodes are increased proportionally. A total of 10,800 4K$\times$4K
image tiles are used for the runs on 100 nodes. 

\begin{figure}[htb!]
\begin{center}
\includegraphics[width=0.95\textwidth]{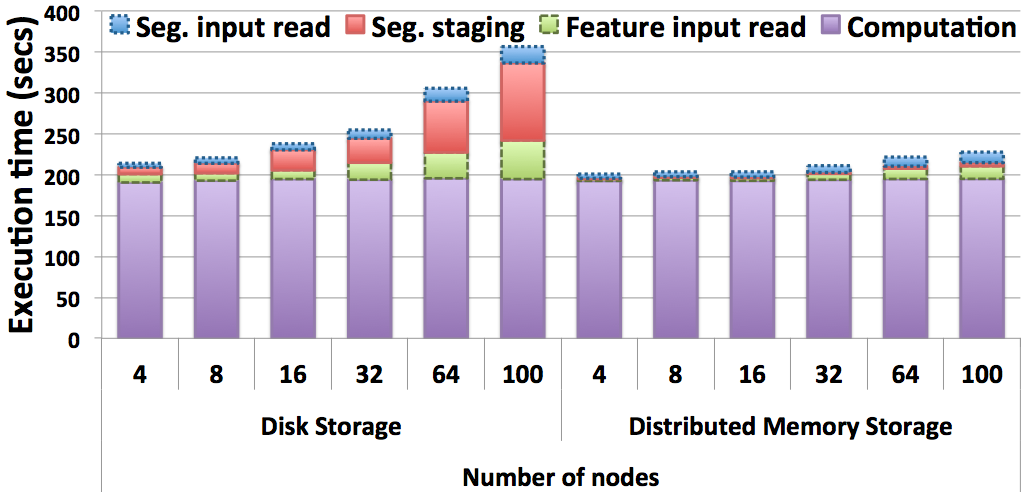}
\vspace*{-1ex}
\caption{Efficiency and scalability of distributed memory
storage and high performance disk storage implementations of
Region Templates. In this evaluation, Region Templates is used to transfer data
from Segmentation to Feature Extraction stages of our example application.
As presented, the cost of the data transfers
with the use of disk storage increases quickly as the number of nodes used grows,
whereas the distributed memory storage based mechanism attains better
efficiency and scalability.}
\vspace*{-2ex}
\label{fig:ds-vs-fs}
\end{center}
\end{figure}

The performance of the example application using the DISK and DMS
based implementations is presented in Figure~\ref{fig:ds-vs-fs}. As
shown, the DMS version achieved better performance in the baseline
configuration (4 nodes) and higher scalability when the number of nodes is
increased. Figure~\ref{fig:ds-vs-fs} shows that the cost of writing the output 
of the segmentation phase (``Seg. staging'') using DMS is at least
10$\times$ smaller than that using the DISK based version. We should note that
the DMS version writes the ``Mask'' and ``RGB'' data regions in this
stage, while the DISK version writes out the ``Mask'' only, since the
``RGB'' data region is already stored in the file system. As a consequence, the
amount of data staged in by the DMS version is 4$\times$ of the amount
of data staged in by the DISK version. 

Although the DISK and DMS versions of the application read the
input data regions for the segmentation stage (``Seg. input read'') from the
file system, the cost of this operation is cheaper in the DMS based
executions. This better performance is a side effect of the DMS 
version not using the file system to exchange data between the segmentation and
feature computation stages, which leads to lower load on the file system, and
hence to less expensive reading operations as compared to the DISK 
version.

\begin{figure}[htb!]
\begin{center}
\includegraphics[width=0.95\textwidth]{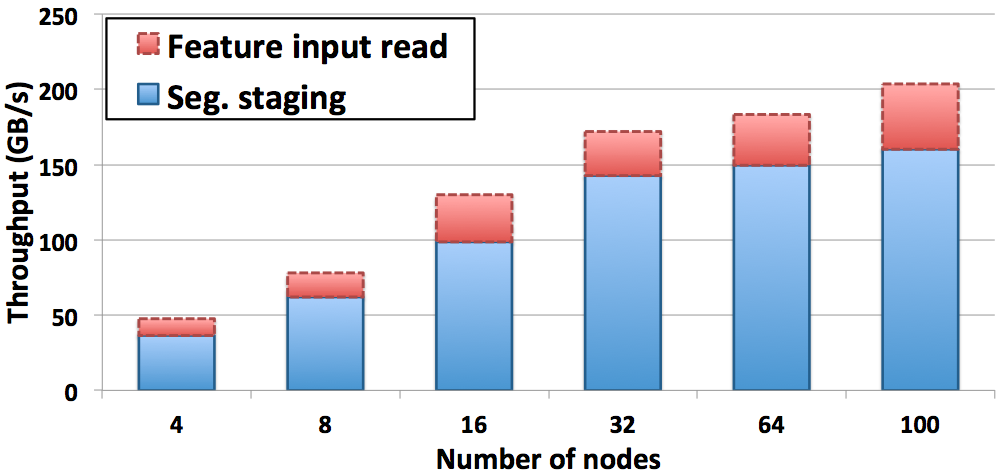}
\vspace*{-1ex}
\caption{Data staging/reading throughput of the distributed memory storage (DMS)
based global region templates implementation (built on top of DataSpaces) for
the communication among the Segmentation and Feature Computation stages of the
example application.}
\vspace*{-2ex}
\label{fig:ds-throughput}
\end{center}
\end{figure}

Further, the data transfers rates (GB/s) among the application stages for the
DMS implementation are presented in
Figure~\ref{fig:ds-throughput}. The region templates achieves very high
communication throughput among stages, reaching an aggregate transfer rate of
about 200~GB/s. The process of reading the data regions for the feature
computation stage (``Feature input read'') using the DMS version is
about 3-4$\times$ faster than that using the DISK version. This
performance improvement is a result of a DataSpaces (used by the DMS) implementation design decision that
optimizes data insertion operations. When a process inserts data into
DataSpaces, the data is stored on a single DataSpaces server (node) and only
the metadata of the data is propagated to the other servers in the system.
This scheme avoids data duplication and unnecessary data movement and split.
The read operation, on the other hand, may result in a data movement that
cannot be avoided and is more expensive.
\subsection{Cooperative CPU-GPU Executions}
\begin{figure*}[h!]
\begin{center}
\mbox{
 \subfigure[RT execution time.]{\label{fig:rtCPUexec} 
       \includegraphics[width=0.95\textwidth]{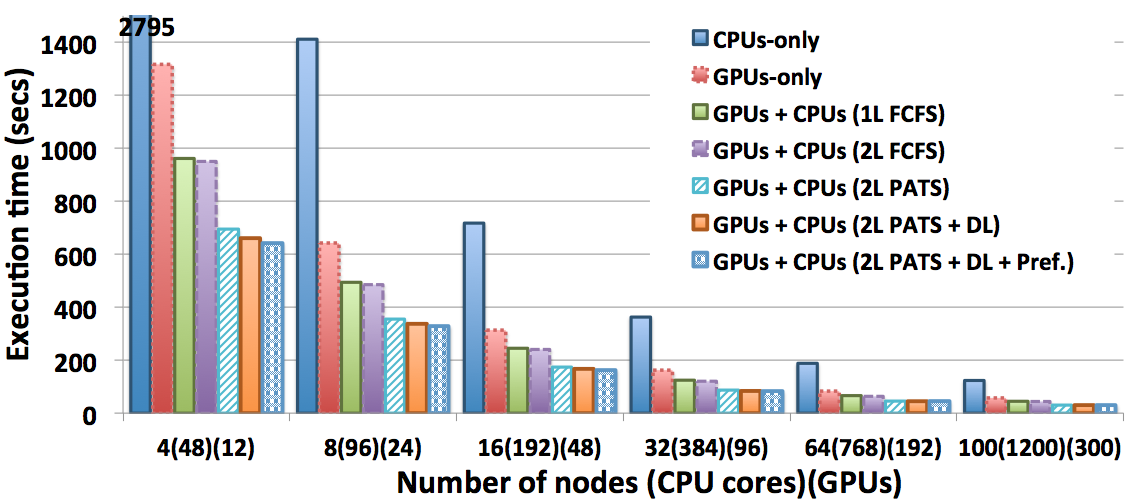}}
}        
\mbox{
 \subfigure[Speedup on top of the multicore CPU-only version.]{\label{fig:rtCPUIOeffi} 
       \includegraphics[width=0.95\textwidth]{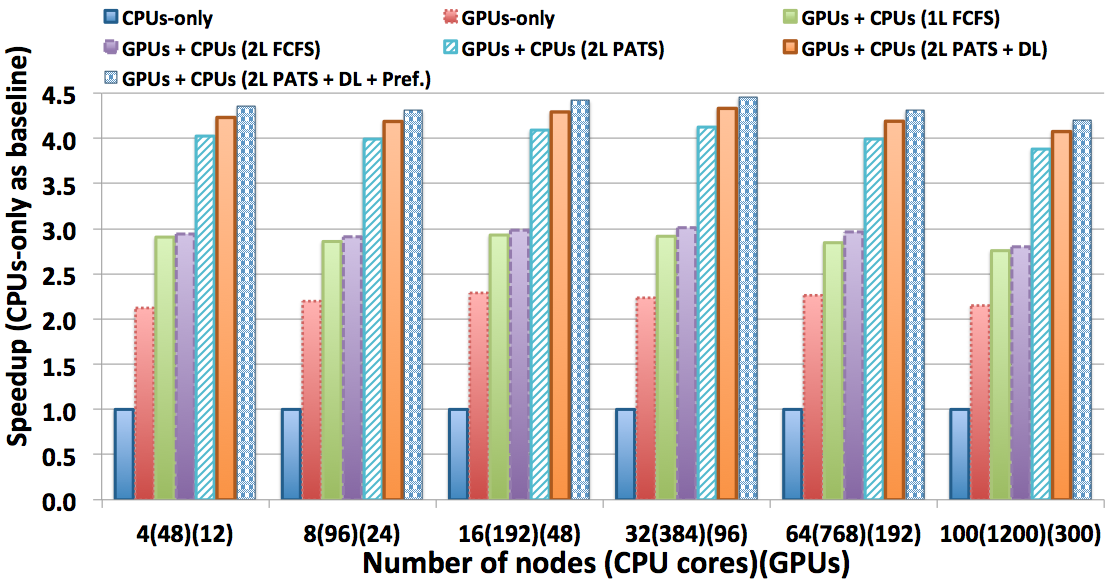}}
}
\vspace*{-1ex}
\caption{Performance of application using multiple versions:
(i)~\emph{CPUs-only} is the CPU multi-core version that uses all the 12 cores
available in each node; (ii)~\emph{GPUs-only} uses only the 3 available GPUs in
each machine; (iii)~\emph{GPUs + CPUs (1L)} uses the CPUs and GPUs in
coordination and the application stages are represented as using a single task
that bundles all the internal operations; (iii)~\emph{GPUs + CPUs (2L)}
utilizes CPUs and GPUs in coordination and represents the application as a
hierarchical computation graph with two levels. FCFS and PATS scheduling
strategies are used in cooperative executions, as well as locality conscious
tasks assignment (DL) and data prefetching and asynchronous data copy (Pref.)
optimizations are employed with the best cooperative version of the
application. All configurations attain good scalability and cooperative
executions resulted in significant improvements on CPUs-only or GPUs-only
versions. Further, the 2L configuration with PATS achieved better performance
than FCFS, because of its ability to exploit variability in speedups of
individual tasks of the application (Figure~\ref{fig:speedup-break}) to
optimize utilization of heterogeneous devices. Additionally, the combined use
of DL and Pref. optimization resulted in a speedup of 1.08$\times$.}
\vspace*{-2ex}
\label{fig:CPU-GPU-RT}
\end{center}
\end{figure*}

In these experiments, we used a fixed set of 104 images (for a total of 6,212
4K$\times$4K tiles) as the number of nodes is varied. Four versions of the
application were executed: (i)~\emph{CPUs-only} is the CPU multi-core version
that uses all the 12 cores available in each node; (ii)~\emph{GPUs-only} uses
only the 3 available GPUs in each machine for computation; (iii)~\emph{GPUs +
CPUs (1L)} uses the CPUs and GPUs in coordination, but the application stages
are represented as a single task that bundles all the internal
operations; (iii)~\emph{GPUs + CPUs (2L)} utilizes CPUs and GPUs in
coordination and represents the application as a hierarchical computation graph
with two levels.  Therefore, the fine grain operations in a stage are
dispatched for computation as individual tasks. Both cooperative CPU and GPU
versions (1L and 2L) can employ FCFS (First-Come, First-Served) or the PATS
scheduling strategy. In addition, we also evaluate the performance benefits of
employing data locality conscious tasks assignment (DL) and data prefetching
and asynchronous data copy (Pref.) with the application version that attained
best execution times.

\begin{figure}[h!]
\begin{center}
\includegraphics[width=0.85\textwidth]{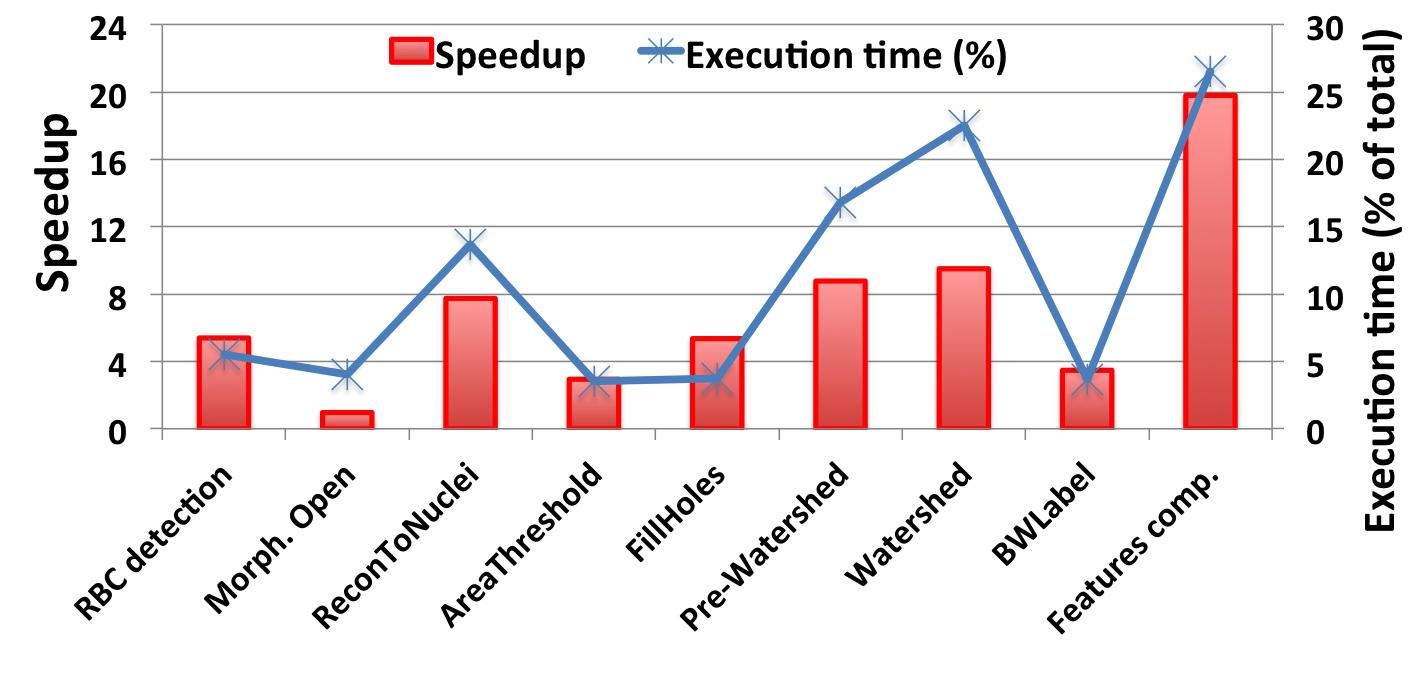}
\vspace*{-1ex}
\caption{GPU speedup values for each function in the segmentation and feature
computation stages. A strong variation on speedups values is observed due
differences on individual operations computation and data access patterns,
which make them more appropriate for different computing devices. PATS exploits
this variation to better use available devices.}
\vspace*{-2ex}
\label{fig:speedup-break}
\end{center}
\end{figure}

\emph{Scalability and schedulers evaluation.} The execution times for the various versions of the application are presented
in Figure~\ref{fig:rtCPUexec}. All versions achieved good scalability -- the
efficiency of the CPUs-only version using 100 nodes was 90\%. The GPUs-only
version attained a speedup of about 2.25$\times$ on top of the CPUs-only, as
shown in Figure~\ref{fig:rtCPUIOeffi}.  The cooperative CPU-GPU execution using
a single level computation graph (GPUs + CPUs (1L)) and FCFS to distribute
tasks among CPUs and GPUs attained a speedup of 2.9$\times$ on the CPUs-only
version.  Figure~\ref{fig:rtCPUexec} also presents the results for the
hierarchical version of the application (2L). As shown, the 2L configuration
with the PATS scheduler was able to significantly (about 1.38$\times$) improve
any other configuration that employs cooperative CPU-GPU execution. In
addition, ``GPUs + CPUs (2L PATS)'' achieved a performance improvement of near
4$\times$ on top of the multi-core CPUs-only version. The best performance of
PATS with 2L is the result of its ability to assign subtask in a stage to the
most appropriate devices, instead of assigning an entire stage for execution
with a single processor as in the 1L configuration.
Figure~\ref{fig:speedup-break} presents the GPU speedups of the individual
operations in each stage. We observe that there is a strong variation in the
amount of acceleration among the functions, because of their different
computation patterns.  This variation in functions attained acceleration, as discussed,
is exploited by the performance-aware task scheduler (PATS) to attain better
performance.

\emph{Performance of data movement optimizations.} Further, we evaluate the
performance benefits of data locality conscious tasks assignment (DL) and data
prefetching and asynchronous data copy (Pref.) with the 2L PATS version of the
application. As presented in Figure~\ref{fig:rtCPUexec}, the use of DL improves
the application performance in 1.05$\times$, because of the reduction in data
transfer volume attained by the scheduler with this optimization.  Finally,
the Pref. optimization is used with the previous version of the application,
and a speedup of 1.03$\times$ is achieved. For this use case application, the
data transfers times represent 12\% of the total execution time and, as
presented, we attained combined improvement on the application execution time
of about 8\% (1.08$\times$). Therefore, our optimization techniques were able
to eliminate near 66\% of the data transfer costs, while we still able to
employ efficient scheduling to optimize utilization of hybrid processors.
Finally, the ``GPUs + CPUs (2L PATS + DL + Pref.)'' version of the application
achieved the best performance with a speedup of 4.34$\times$ on top of the
multicore CPU-only version, as presented in Figure~\ref{fig:rtCPUIOeffi}.

\begin{figure}[h!]
\begin{center}
\mbox{
 \subfigure[Impact of inaccurate speedup estimations.]{\label{fig:error-speedup} 
       \includegraphics[width=0.65\textwidth]{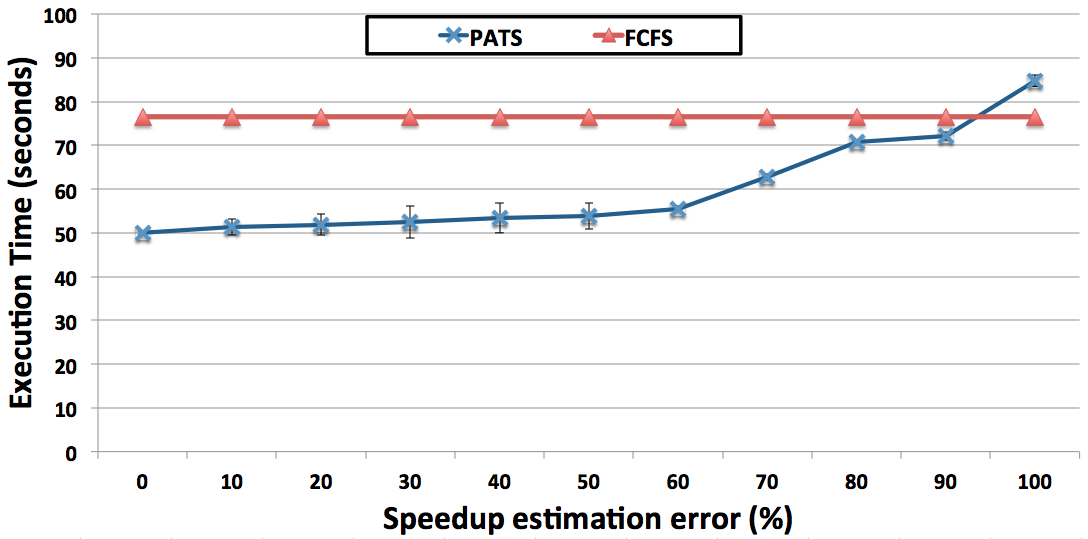}}
}        
\mbox{
 \subfigure[PATS profile: \% tasks executed by GPU vs error rate.]{\label{fig:erro-rate-prof} 
       \includegraphics[width=0.65\textwidth]{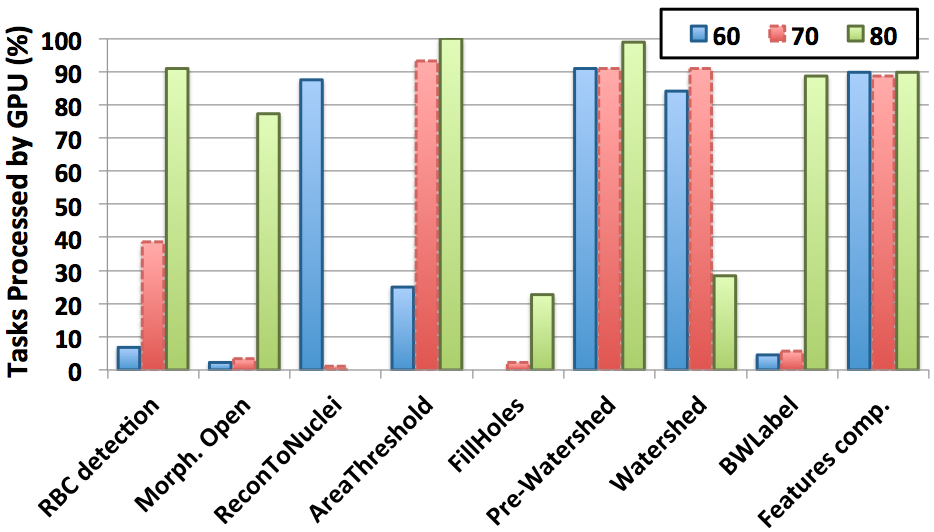}}
}
\vspace*{-1ex}
\caption{Evaluation of the PATS performance under different levels of
inaccurate speedup estimates. To effectively confound the method and force
wrong assignment of operations to processors, tasks with low speedups have their
speedups increased by a given percentage, whereas tasks with high speedups have
their estimates decreased. As shown, PATS performs well even with high speedup
estimation error and, for instance, it suffers a performance degradation of only
8\% when a percentage error of 50\% is used. In addition, to understand 
how very high estimate errors affected the scheduling, we present the profile of
tasks executed by the GPU as the error rate is varied from 60\% to 80\%, which
refers to the first interval in which there is a significant performance
degradation in PATS. }
\vspace*{-2ex}
\label{fig:inacurrate-est}
\end{center}
\end{figure}

\emph{Sensitivity to inaccurate speedup estimation.} We have also evaluated the
impact of errors in speedup estimates to PATS performance. Since PATS relies on
speedups to keep a sorted queue of tasks, we inserted errors by increasing
tasks with lower speedup (RBC detection, Morph. Open, AreaThreshold, FillHoles,
and BWLabel) and decreasing estimates for other tasks that have higher
speedups.  Our intention was to enforce that at a certain error level, tasks
that are executed with a GPU without estimate errors would change positions
with tasks that otherwise would be computed by a CPU using PATS sorted queue.  The
execution times of the application using PATS and FCFS schedulers, as the error
inserted varies from 10\% to 100\%, are shown in
Figure~\ref{fig:error-speedup}. As presented, the PATS is impacted by
inaccurate speedup estimates, but performance changes are small until high
levels of errors are inserted. For instance, if the error level is 50\%, the
overall performance of the application with PATS decreases only by 8\% as
compared to the configuration in which no error is inserted.  In addition, PATS
is only outperformed by FCFS in the configuration in which 100\% of error is
used, which in practice consists in an inversion of tasks in the sorted queue.
In this configuration, CPU executes tasks that in have high speedup, whereas
GPU processes tasks that with no error estimates would have the smaller
speedups. In order to better understand performance of PATS scheduling with
high percentage of estimate errors, we have collected the profile of the tasks
executed by a GPU when errors of 60\%, 70\%, and 80\% are used.
Figure~\ref{fig:erro-rate-prof} presents the percentage of tasks computed by
the GPU in each configuration. The results show a significant increasing in the
number of tasks with low speedups executed by GPU as error grows. In addition,
other tasks such as "ReconToNuclei" and "Watershed" follow an inverse trend and
are assigned in a smaller number for GPU computation as error level is
increase. Still, even with 80\% of estimate speedup error PATS is faster than
FCFS, since some operations such as "Feature Computation" are correctly
dispatched for execution with GPU.

\section{Related Work} \label{sec:related}
%
%
The Region Templates framework leverages data description concepts proposed by
Baden et al.~\cite{Kohn:1995:PSI:224170.224283}, as we allow for hierarchical
representation of a low dimensional data domain as used in adaptive mesh
methods. Other systems, such as Fortran D~\cite{Fox90fortrand} and Vienna
Fortran~\cite{cs2079}, propose frameworks for parallel execution of arrays of
data elements associated with application specific phenomena. Recent projects
have developed data management systems for array based multi-dimensional
scientific
datasets~\cite{Brown:2010:OSL:1807167.1807271,Tran:2012:TSA:2146382.2146387,6092339}.
SciDB~\cite{Brown:2010:OSL:1807167.1807271} implements several optimized query
operations such as slicing, sub-sampling, and filtering to manipulate large
datasets, as well as it is performs multi-dimensional aware data striping and
versioning. The Pyramid approach~\cite{Tran:2012:TSA:2146382.2146387} includes
similar data types, but is optimized for scalability of metadata management. It
also introduces an array-oriented data access model with active storage
support, which is very useful, for instance, to execute filtering operations
or/and data aggregation closer to or in the data sources. Our framework differs from these
systems in several ways. It enables association of data from multiple sources
targeting the same spatial region, which is a common scenario in sensor data
analysis where multiple data measurements may be taken for the same region,
e.g., measurements of the humidity of certain region over time in monitoring
and change detection analysis~\cite{Chandola:2011:SGP:2024035.2024041}. It
provides a framework for automated management of data across several memory
layers on a parallel system, including multiple implementations for efficient
I/O, while providing a well-defined interface for data retrieval and storage.
Also, different data types commonly used in microscopy image data are
supported, as well as it incorporates the data management with a runtime system
for efficient execution of dataflow application. Because of the increasing
power of hybrid systems equipped with accelerators, the support for data
structures with implementations for CPUs and GPUs and scheduling techniques for
hybrid systems are other important feature included in region templates. 

Efficient execution of applications on CPU-GPU equipped platforms has been an
objective of several projects in the past
years~\cite{mars,merge,qilin09luk,1616864,Diamos:2008:HEM:1383422.1383447.Harmony,6061070,ravi2010compiler,6152715,hpdc10george,6267858,Rossbach:2011:POS:2043556.2043579,Teodoro:2012:Cluster,augonnet:hal-00725477,cluster09george,HartleySC10,hartley,saltz2013feature,teodoro2013comparative,teodoro-2013-vldb}.
The EAVL system~\cite{DBLP:conf/egpgv/MeredithAPS12} is designed to take
advantage of CPUs and GPUs for visualization of mesh based simulations.  Ravi
et al.~\cite{ravi2010compiler,6152715} propose techniques for automatic
translation of generalized reductions. The OmpSs~\cite{6267858} supports
efficient asynchronous execution of dataflow applications automatically
generated by compiler techniques from annotated code.   DAGuE~\cite{6061070}
and StarPU~\cite{1616864,augonnet:hal-00725477} are motivated by regular linear
algebra applications on CPU-GPU machines.  Applications in these frameworks are
represented as a DAG of operations, and different scheduling policies are used
for CPU and GPU task assignments. These solutions assume that the computation
graphs are static and known prior to execution. This is a serious limitation to
dynamic application such as ours, as the execution of the next stage may be
dependent on a computation result. Our framework allows for the dependency
graph to be built at runtime.

KAAPI~\cite{Gautier:2007:KTS:1278177.1278182} is a programming framework the
supports execution of dataflow applications on distributed memory systems. It
is inspired on the Athapascan-1~\cite{727176} programming language that allows
for a precise description of dependencies in dataflow graphs, and includes
asynchronously spawned tasks with explicit shared variable to facilitate data
dependency detection. The mapping of tasks onto processors is carried out dynamically employing work-stealing techniques.
In~\cite{Hermann:2010:MMP:1885276.1885302}, KAAPI was used and extended to
parallelize an iterative physics simulations application in machines equipped
with GPUs and CPUs. This work proposed novel scheduling strategies for
dataflows in hybrid systems, which includes runtime load balance with initial
workload partition that takes into consideration affinity to objects accessed.
XKaapi~\cite{conf/ipps/GautierLMR13} further extended KAAPI with support for
execution on hybrid systems, equipped with CPUs and GPUs. In XKaapi, the
programmer develops applications using a multi-versioning scheme in which a
processing task may have multiple implementations targeting different
computing devices. In order to efficiently execute dataflows in heterogeneous
system, it implements new scheduling strategies and optimizations for reducing
impact of data transfers between devices, locality-aware work stealing, etc.
XKaapi is also evaluated using regular linear algebra applications. More
recently, XKaapi was deployed in new Intel Xeon Phi
coprocessor~\cite{DBLP:conf/sbac-pad/LimaBGR13}.  XKaapi is contemporary to our
runtime systems and, as such, they share a number of optimizations, including
strategies for reducing impact of data transfers, locality-aware scheduling,
etc. In special, the performance-aware scheduling strategy proposed in our work
is not available in other systems, whereas we plan to incorporate the ideas of
KAAPI application description to perform implicit calculation of data
dependencies in dataflow graphs of region templates applications.

Systems such as Linda~\cite{Carriero:1989:LC:63334.63337},
ThreadMarks~\cite{Amza:1996:TSM:226705.226708} and Global
Arrays~\cite{nieplocha97} were proposed to provide shared-memory programming
model abstractions -- typically using matrices to represent data -- on
distributed platforms with Non-Uniform Memory Access (NUMA).  These systems
have been very successful, because they simplify the deployment of applications
on distributed memory machines and are efficient for applications that exhibit
high data locality and access to independent data blocks of a matrix. Data
access costs in these systems tend to increase for patterns involving frequent
reads and writes of data elements that are not locally stored on a processor.
The efficiency and applicability of these solutions is hindered in such cases.
Like these systems, our framework provides an abstraction for the
representation and computation of large volumes of data on distributed memory
machines. However, we do not intend to provide a generic distribute shared
memory in which several processes have access to the entire data domain and
consistency is taken care by the system. In our case, data accessed are well
defined in terms of input region templates, which allows for data to be moved
ahead of computation and restricts data access to locally loaded data.

FlexIO~\cite{6569822} is a middleware that supports flexible data movement
between simulation and analytic components in large-scale computing systems.
It implements an interface for passing data, originally proposed in
ADIOS~\cite{LofsteadKSPJ08}, that mimics "files" manipulations.  Data are
exchanged during I/O timesteps of simulation applications via either disk or
memory-to-memory mechanisms. FlexIO supports several placement architectures,
including (i)~"Simulation core": I/O is performed by the application computing
cores; (ii)~"Helper Core": cores on the same node receive and stage data from
computing cores; (iii)~"Staging Core": I/O cores are placed in separate nodes;
(iv)~"Offline": output simulation data is moved to disk for further analysis.
Is also proposes automated heuristics to find the appropriate I/O placement of
applications. Placement is static and remains the same during execution.  Such
as FlexIO, the global data storage module of region template supports multiple
strategies for placement computing and I/O cores, which may be easily
interchangeable by the application developer. However, region templates
provides a richer set of data abstraction with data types such as sparse and
regular arrays, matrices, and objects, which are managed by the system and may
be staged to different storage mechanisms. ADIOS/FlexIO use a "file" like
interface to write/read data in which variables are associated with data
elements (scalar, arrays, etc). Multiple processes that manipulate the same
file need to deal with low-level details, such as data packing, data offsets
in global domains, process group data distribution and partition, etc, whereas
it is abstracted in region templates.  In addition, as discussed in
Section~\ref{sec:storage-impl}, our implementation of global storage is
optimized for asynchronous applications, which differently from simulation
applications do not have synchronization points in which all processes
synchronize to perform I/O operations. As presented in our experimental
results, the use of flexible I/O groups lead to improved performance on our use
case application as compared to the approach originally implemented in ADIOS
for iterative simulations. Finally, as a future work we want to provide
automated placement on distributed system, and the propositions of FlexIO may
be integrated and optimized for our scenario.

\section{Conclusions and Future Work} \label{sec:conc}
Researchers have an increasing array of imaging technologies to record detailed
pictures of disease morphology at the sub-cellular levels, opening up new
potentials for investigating disease mechanisms and improving health care
delivery. Large clusters of GPU equipped systems, in which each {\em hybrid}
node contains multiple CPUs and multiple GPUs, have the memory capacity and
processing power to enable large imaging studies. However, these systems are
generally difficult to program due complexities arising from the heterogeneity
of computation devices and multiple levels of memory/storage hierarchies (going
from persistent disk-based storage on a parallel file system to distributed
memory on the cluster to memories on CPUs and GPUs within a node). 

The region template abstraction aims to hide the complexity of
managing data across and within memory hierarchies on hybrid systems for
microscopy image analysis applications. Some of the characteristics that allow
for the efficient execution and data management of region templates are the
following: (1)~region template applications are instantiated as a
graph of computation stages and communication only exists among different
stages of the application. Therefore, computation within a given stage uses
exclusively local data received as input to the stage; (2)~data chunks
accessed within a given stage instance are exported to the runtime system,
because they are accessed via the region templates interface.  Therefore, the
system knows in advance which data regions (or blocks of a data region) are 
accessed by a stage and can retrieve the data asynchronously, reducing the impact of
data transfer costs; (3)~the mapping of copies of the pipeline stages to
computing nodes can be carried by the runtime system in a way to minimize data
transfers. 

The region template framework provides implementations for common data
structures used in target applications; therefore, expect small overhead when
developing and integrating new applications.

Our experimental evaluation shows that very high processing and data transfer
rates can be achieved in our framework. The processing rate with cooperative
CPU-GPU executions using the 2L PATS configuration and 100 nodes is 11,730
4K$\times$4K tiles (about 117 whole image slides, 750GB of data) per minute.
The combined data staging and reading rates between the computing stages are
about 200GB/s when distributed memory storage is used.  This level of
performance will enable much larger imaging studies and is a very promising
direction that should lead to better understanding of disease behavior. 

We are currently deploying a new biomedical image analysis application on top
of region templates. This application computes large-scale cell tracking to
study of early stages of metastasis in cancer. The goal of the application is
to correlate cell tracking information with other data sources, such as genetic
information, in order to better understand the disease. Beyond the great
potential science results, this application also brings a new challenging
computational scenario. In object tracking, the application only accesses
subsets of the data domain that are likely to contain objects of interest. In
addition, the path followed by an object until the current timestamp tend to be
very useful in identifying in which sub-domain it will be located in future.
This context is motivating extensions in region templates to include smart
spatial-temporal caching and data prefetching strategies, which could, for
instance, anticipate the data reading process and reduce the impact of these
operations to the application.\\

\noindent {\bf Acknowledgments}. This work was supported in part by
HHSN261200800001E from the NCI, R24HL085343 from the NHLBI, R01LM011119-01 and
R01LM009239 from the NLM, RC4MD005964 from the NIH, PHS UL1RR025008 from the
NIH CTSA, and CNPq (including projects 151346/2013-5 and 313931/2013-5). This
research used resources of the Keeneland Computing Facility at the Georgia
Institute of Technology, which is supported by the NSF under Contract
OCI-0910735. 

\clearpage



%
\bibliographystyle{abbrv}
\bibliography{george}  
%

\end{document}